\documentclass[12pt,letterpaper]{article}     
\usepackage[utf8]{inputenc}                   
\usepackage{cite}
\usepackage{geometry}
\usepackage{amsmath}      
\usepackage{amsfonts}     
\usepackage{amssymb}
\usepackage{graphicx}     
\usepackage{rotating}
\usepackage{pdfpages}
\usepackage[lofdepth,lotdepth]{subfig}	
\usepackage{unitsdef}	  
\usepackage[colorlinks=true,urlcolor=blue,linkcolor=black,citecolor=green]{hyperref}     
\usepackage{float}		
\usepackage{booktabs}	
\batchmode
\usepackage{lastpage}
\usepackage{fancyhdr}	
\author{Adrián Francisco Eduarte-Rojas, Francisco Frutos-Alfaro,\\ Rodrigo Carboni, Daniel Alvarado}
\title{Chaotic Behavior of Geodesics in Kerr-like Spacetime}

\begin{document}	

\pdfbookmark[1]{Portada}{portada} 	

\maketitle							


\begin{abstract}

In this contribution, the motion of unitary mass test particles in a perturbed Kerr-like metric is studied using simulations in the configuration and phase space. Our metric represents the approximate exterior spacetime of a massive rotating body with mass quadrupole moment $q$. Chaotic behavior arises as the $q$ parameter increases. The rotation number is determined for an axis of symmetry in the corresponding Poincaré section. The existence of chaotic regions in the region near the source event horizon is shown.

\end{abstract}


\begin{small}
\end{small}

\tableofcontents

\section{Introduction}

The recent advances in the direct observation of the motion of stars orbiting around a compact massive object, like black holes, have made possible to compare among different spacetime solutions of the Einstein Field Equations (EFE). One of the most important solutions of EFE in the study of gravity in compact object is the Kerr metric which models the exterior spacetime of a rotating source. In order to find the path of a particle, we only need the information of the source mass and its spin parameter or intrinsic angular momentum and the initial conditions of the particle \cite{gravitationmisnerthorne}. The equations of motion of the Kerr metric are completely integrable given that there are four constants of motion: rest energy of the particle or its invariant mass $\mu$, the $z$-component of the angular momentum $L_z$, the particle energy $E$ and the Carter constant $\mathfrak{C}$ \cite{gravitationmisnerthorne}. However, the conditions for a real compact object to fit into the Kerr model are too specific, a real object like earth or even the sun have a slightly deviation from a perfect sphere because of the rotation and other factors. If the gravity source is flattened then the Kerr metric is no longer valid and thus it is expected the absence of the Carter constant in the equations of motion. Moreover, this equations could be non integrable in the analytic sense and the evolution of the particle's motion could be chaotic under certain circumstances.

The detection of gravitational waves could be evidence of the possible chaotic motion of particles around the gravity source. The gravitational waves spectra is highly related with the phase space of the particle moving around the compact object and thus if the source breaks the spherical symmetry then the chaos will be reflected in the phase space \cite{ChaoticShadow, GrowthResonances, TestingSpacetime}. In order to study the geodesics produced by an strong gravitational field, we use a perturbed EFE solution known as Kerr-like metric with mass quadrupole moment \cite{frutosmetric, frutos2}. This metric represents the spacetime surrounding a massive, deformed and rotating compact object; like neutron stars, among others. The quadrupole mass moment $q$ measures the deviation from a spherical gravitational source. Moreover, when $q = 0$ the Kerr-like metric and the original Kerr metric match. Therefore, the gravitational source may be described by its rotation is represented by the spin parameter $a$, its mass by $M$ and the mass quadrupole moment $q$ \cite{frutosmetric, frutos2, gravitationmisnerthorne, weinberg}.

\section{The Kerr-like Metric and Equations of Motion}

In general, the spacetime surrounding a rotating compact object has a metric with the following structure

\begin{equation}\label{eq:mgen}
ds^2 = - Vdt^2 + 2Wdtd\phi +Xdr^2 +Yd\theta^2 +Zd\phi^2 
\end{equation}

\noindent
where the potentials $V$, $W$, $X$, $Y$, $Z$ are functions of $r$ and $\theta$ only. 

For the Kerr-like metric with mass quadrupole moment, the potentials take the following form \cite{frutosmetric, frutos2}:

\begin{eqnarray}
V &=& \frac{e^{-2\psi}}{\rho^2} (    \Delta -a^2 \sin ^2 \theta )   \nonumber \\
W &=& -\frac{2Jr}{\rho^2}\sin^2 \theta    \nonumber \\
X &=& \rho^2 \frac{e^{2\chi}}{\Delta}   \\
Y &=& \rho^2 e^{2\chi}    \nonumber \\
Z &=&  \frac{e^{2\chi}}{\rho^2}( (r^2 +a^2)^2 - a^2\Delta \sin ^2\theta   )\sin ^2\theta  \nonumber 
\end{eqnarray}

\noindent
where
 
\begin{equation}
\psi = \frac{q}{r^3}P_2 + 3 \frac{Mq}{r^4}P_2
\end{equation}

\begin{equation}
\chi = \frac{qP_2}{r^3} + \frac{1}{3}\frac{Mq}{r^4}  (-1 + 5 P_2 + 5 P_2 ^2) + \frac{1}{9} \frac{q^2 }{r^6}(2 + 6 P_2 + 21 P_2 ^2 + 25 P_2 ^3) 
\end{equation}

\noindent
and $ P_2 $ is the Legendre polynomial

\begin{equation}
P_2 = \frac{1}{2} {(3\cos ^2 \theta  -1)} 
\end{equation}

Although we can use the geodesic equations to easily get the four second order equations of motion, in order to apply the dynamical system theory only first order equations of the form $\dot{x} = f(x)$ are needed. Therefore, the langragian and hamiltonian formalisms are used to find this kind of equations of motion. By defining the lagrangian as $\mathcal{L} = {\mu}/{2}({ds}/{d\tau})^2$, where $\tau$ is an affine parameter. The four-momenta $p_\mu = {\partial \mathcal{L}}/{\partial \dot{x}^\mu}$ could be easily obtained

\begin{equation}
\begin{split}
p_t &= \mu ( -V \dot{t} +W\dot{\phi})= -E \\
p_r &= \mu X \dot{r}   \\
p_\theta &= \mu Y\dot{\theta}   \\
p_\phi &= \mu (W\dot{t} + Z \dot{\phi}) = L_z
\end{split}
\end{equation}

Note that the lagrangian does not depend directly of $t$ and $\phi$, hence their respective conjugated momenta are actually constants of motion. Now, we turn to the hamiltonian formalism to find the remaining equations of motion. By examining the lagrangian, it is easy to see that the hamiltonian is also a constant of motion related to the rest energy or the invariant mass, remembering that $p_\mu p^\mu = -\mu^2$, then $\mathcal{E} = -\mu$ .

\begin{equation} \label{E:condini}
\mathcal{H} = \frac{\mu}{2} \left(  -\frac{E^2 Z + 2EL_z W -L_z ^2 V}{\mu^2 (VZ+W^2)}  +  \frac{  p_r ^2   }{ \mu^2 X   } +  \frac{  p_\theta ^2   }{ \mu^2 Y   }     \right)  =\mathcal{E}
\end{equation}

Finally, we take the canonical equations to find the dynamical system model

\begin{equation} 
\label{E:ecmov}
    \begin{split}
        \dot{r} &= p_r/\mu X \\
        \dot{\theta} &= p_\theta /\mu Y \\
        \dot{t} &= \frac{1}{ \mu \tilde{\rho}^2}( EZ + L_z W )  \\
        \dot{\phi} &= \frac{1}{\mu  \tilde{\rho}^2}( L_z V - EW ) \\
        \dot{p_r} &= \frac{1}{2} \left(  p_r ^2 \partial_r X /X^2 + p_\theta ^2 \partial_r Y /Y^2 + \partial_r V_{alm} \right) \\
        \dot{p_\theta} &= \frac{1}{2} \left(  p_r ^2 \partial_\theta X /X^2 + p_\theta ^2 \partial_\theta Y /Y^2 + \partial_\theta V_{alm} \right) \\
        \dot{p_\phi} &= 0  \\
        \dot{p_t} &= 0 
    \end{split}
\end{equation}
    
To find the solution, it is important to take into account that only one initial condition for the momenta is needed, because the equation (\ref{E:condini}) constrains the radial momentum $p_r$ with the angular momentum $p_\theta$. To simplify the numerical solutions, we take the particles mass $\mu = 1$ and later select a collection of parameters and initial conditions.

\section{Dynamics Analysis}

In order to study the behaviour of chaotic systems, there are two important theorems that characterize the evolution of the geodesic. Both theorems play a central role in hamiltonian systems perturbed by small parameters. The Kolmogorov-Arnold-Moser theorem (KAM) says that if the bounded motion of an integrable hamiltonian $\mathcal{H}_0$ is perturbed by a small $\Delta \mathcal{H}$ that makes non integrable the total hamiltonian $\mathcal{H}$ = $\mathcal{H}_0$ + $\Delta \mathcal{H}$, and if the conditions are satisfied, the perturbation $\Delta \mathcal{H}$ is small and the frequencies $\omega_i$ of $\mathcal{H}_0$ are immeasurable. Then, the motion will remain bounded to a $N-torus$, except for a negligible set of initial conditions resulting in a serpentine path over the surface energy \cite{goldstein}. 

The Poincaré-Birkhoff (PB) theorem states from the infinitely many periodic orbits on a resonant torus of the integrable system only an even number survive in the perturbed system. Moreover, half of these surviving periodic orbits are stable and the other half unstable \cite{LukesZpoyVoorhees}. 

According to the KAM theorem most of the surviving orbits will be still confined to a torus, although, they will be slightly deformed regarding the original Kerr system. On the other hand, the Poincare-Birkhoff theorem tell us that the torus breaking will give us new kind of structures in the phase space and it will be visible in a Poincaré section.

To find the numerical solutions of the dynamical system, it was developed a program that finds the geodesics followed by a test particle, thus we studied the behavior by sweeping the mass quadrupole moment parameter, starting with the Kerr case ($q = 0$) until we reach a highly deformed source ($q= 0,7$). The Poincaré sections as a method to distinguish between ordered and chaotic orbits is employed.

\subsection{Mass Quadrupole Moment $q = 0$}

As mentioned earlier, the Kerr-like metric with mass quadrupole moment is an approximate solution of the EFE and it is based on the mass quadrupole parameter as the perturbation \cite{frutosmetric}. Actually, when $q = 0$ the Kerr-like metric reduces to the Kerr metric, hence, there will be no structures in the phase space  besides the main island of stability because Kerr system is integrable. 

This is used as a starting point. A set of parameters to describe an orbit in the Kerr metric ($q = 0$) is selected and later, an increase in $q$ is done fixing the remaining parameters to study the changes in the motion produced by the perturbation.  

We take the parameters $\mu = 1$, $M = 1.0$, $a = 0.99$, $E = 0.932516$, $L_z = 1.2$ and $q = 0$ similar to those appearing in \cite{grossman2012harmonic}. 

\begin{figure}[H]
    \centering
    \includegraphics[width=0.9\textwidth]{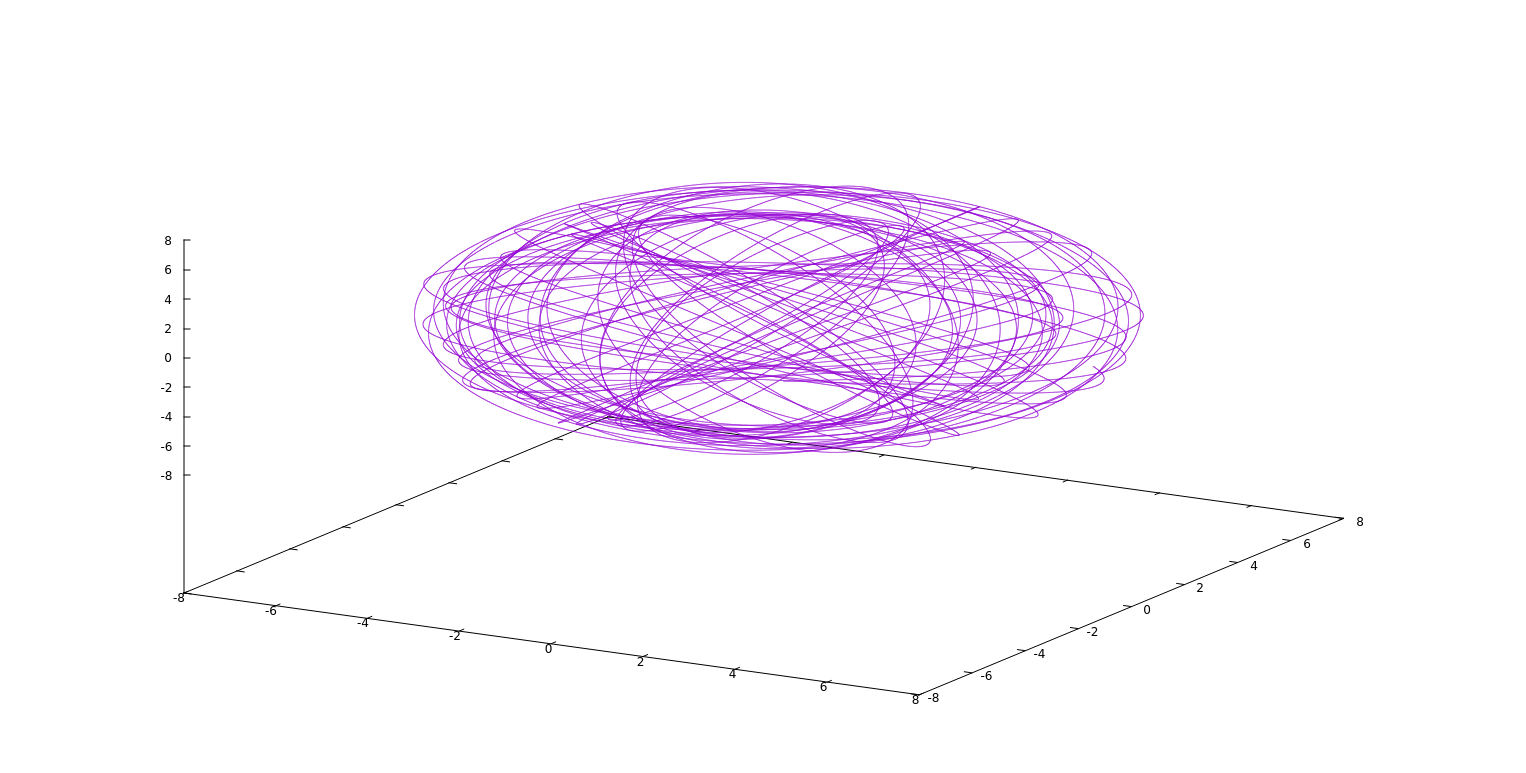}
    \caption{Configuration space of the Kerr metric ($q$ = 0) using the parameters $M = 1.0$, $a = 0.99$, $E = 0.932516$, $L_z = 1.2$ and the initial conditions:  $r = 7.1$, $\theta = \pi/2$, $\phi = 0$, $t = 0$, $p_r = 0$, $p_\phi = L_z = 1.2$, $p_t = -E = -0.932516 $. The initial condition $p_\theta$ was calculated using equation (\ref{E:condini}) and taking the positive square root.}
    \label{F:conf1q0}
\end{figure}

As we can see in figure \ref{F:conf1q0} for this specific set of parameters and initial conditions, the configuration space outside the external event horizon takes a torus shape. Eventually, it will thickly cover the allowed region without falling to the center or escaping to infinity.

\begin{figure}[H]
    \centering
    \includegraphics[width=.9\linewidth]{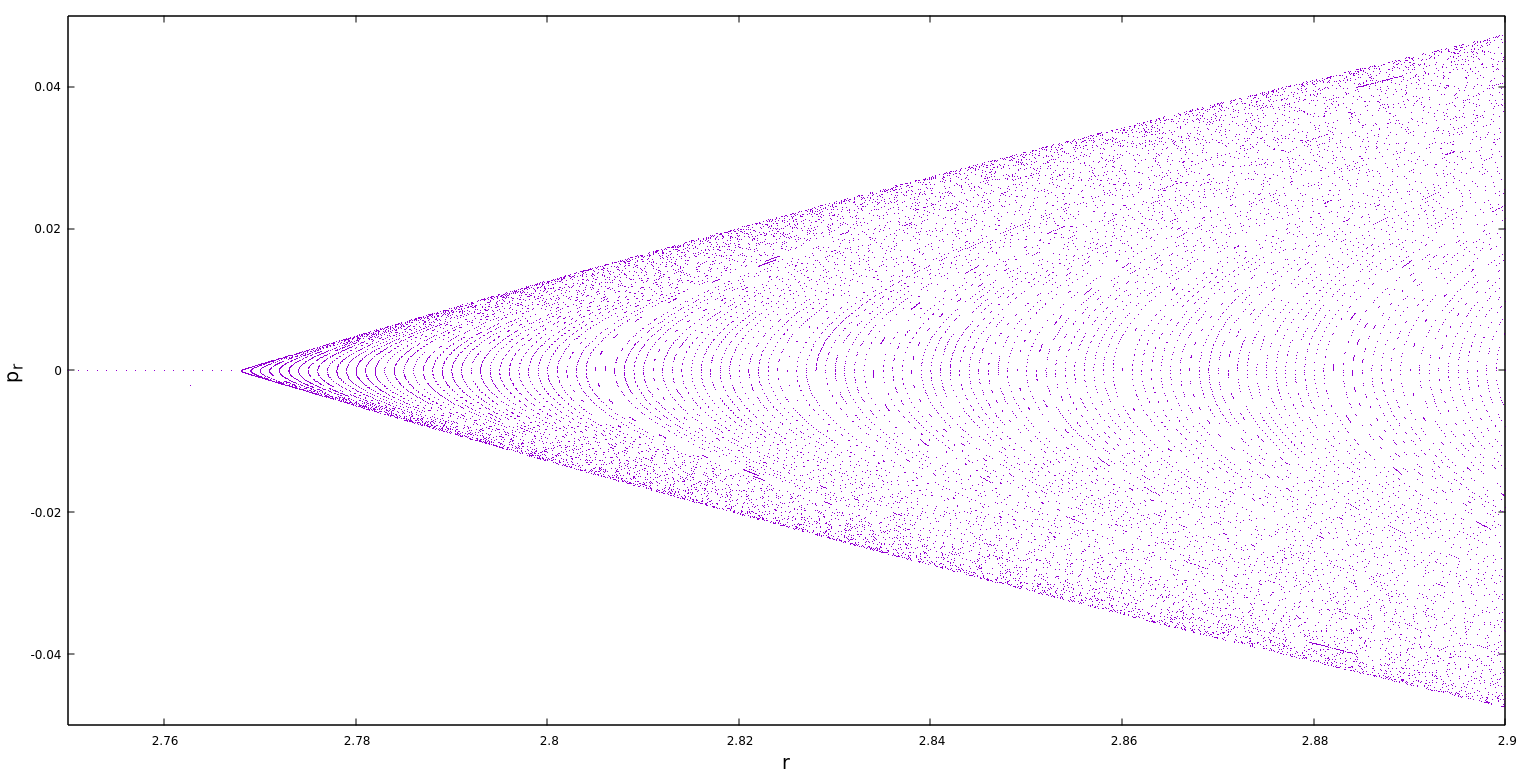}
    \caption{Poincaré section for $\theta = \pi/2$ and $p_\theta \geq 0$ and parameters $M = 1.0$, $a = 0.99$, $E = 0.932516$, $L_z = 1.2$ and $q = 0$, which is exactly a Kerr spacetime system.}
    \label{fig:poincare0a}
\end{figure}

\begin{figure}[H]
    \centering
    \includegraphics[width=0.9\textwidth]{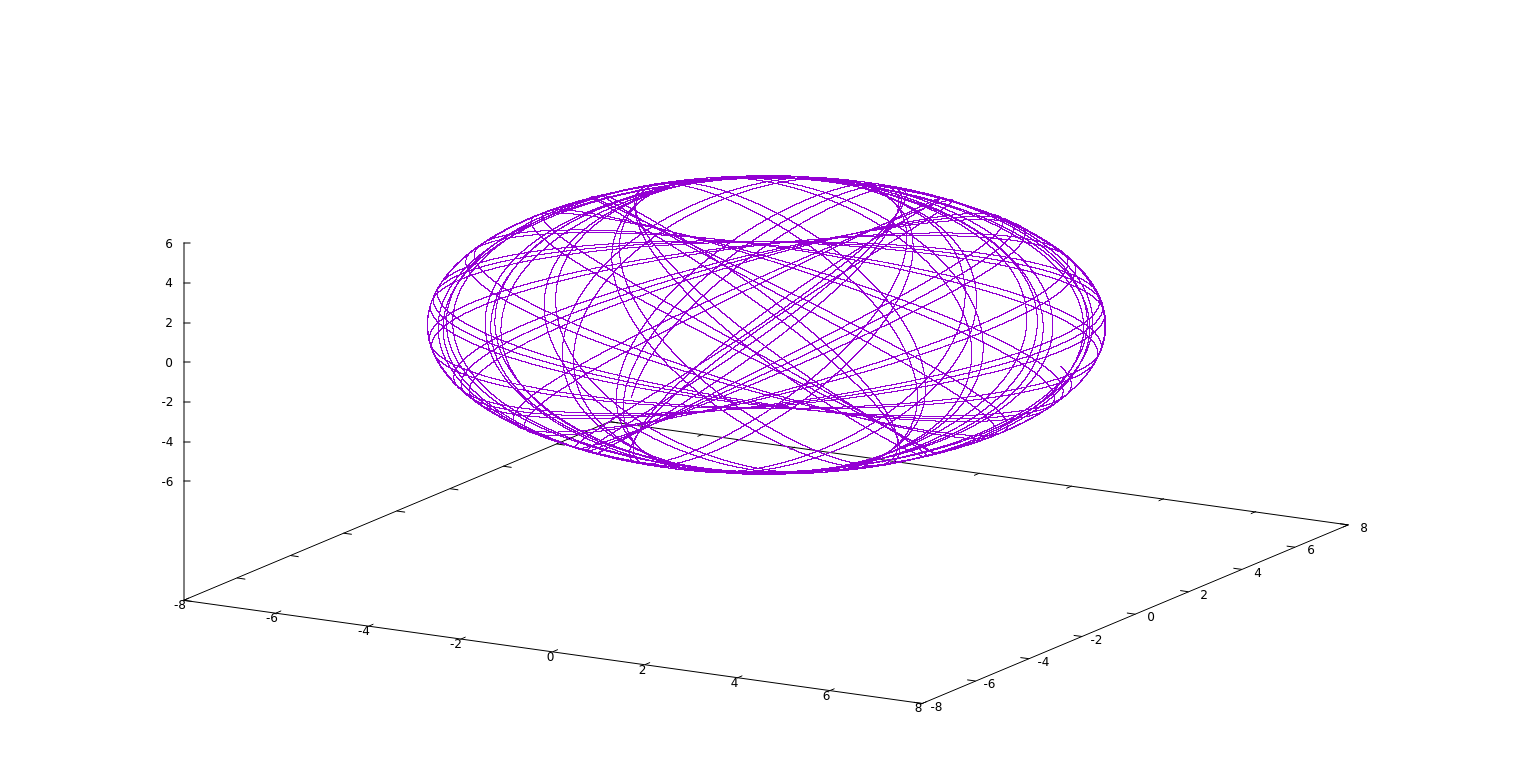}
    \caption{Orbit corresponding to the only elliptic point located at $r = 6.294$, $p_r = 0$ of the Poincaré section of the figure \ref{fig:poincare0a} and for the set of parameters $M = 1$, $a = 0.99$, $L_z = 1.2$ y $E = 0.932516$. The other initial conditions are $\theta = \pi/2$, $\phi = 0$, $t = 0$,  $p_r = 0$, $p_\phi = L_z$ y $p_t = -E$. The initial condition for $p_\theta$ is calculated by equation (\ref{E:condini}).}
    \label{F:q0eliptico}
\end{figure}

The Poincaré section in figure \ref{fig:poincare0a} shows the intersection of the phase space with the plane $\theta = \pi /2$, for all $p_\theta > 0$ and for all $\phi$. In this figure, it is found that the set of all trajectories of the main island of stability form a regular structure of \emph{ellipses}. This is the expected behavior due to the Kerr metric are integrable \cite{ grossman2012harmonic, gravitationmisnerthorne}. Therefore, it is not expected other kinds of structures even near the event horizon \cite{BrinkResonant}. The only elliptical point for this Poincaré section corresponds to $ r = 6.294455$. In figure \ref{F:q0eliptico}, it is shown the corresponding configuration space orbit which is a extremely ordered geodesic because the elliptic point is also a stationary one, therefore $p_r$ is constant.

If the dynamical system is perturbed by the mass quadrupole moment $q$, then some of the torus will break giving new structures like higher order islands, hyperbolic points and chaotic regions. If a small perturbation affects an integrable orbit, then most of the tori will survive deformed and some of the broken tori will show new structures and incipient chaotic behavior \cite{LukesZpoyVoorhees} due the high gravity and the loss of spherical symmetry.

\subsection{Mass Quadrupole Moment $q = 0.3$}  

\begin{figure}[H]
    \centering
    \includegraphics[width=0.9\textwidth]{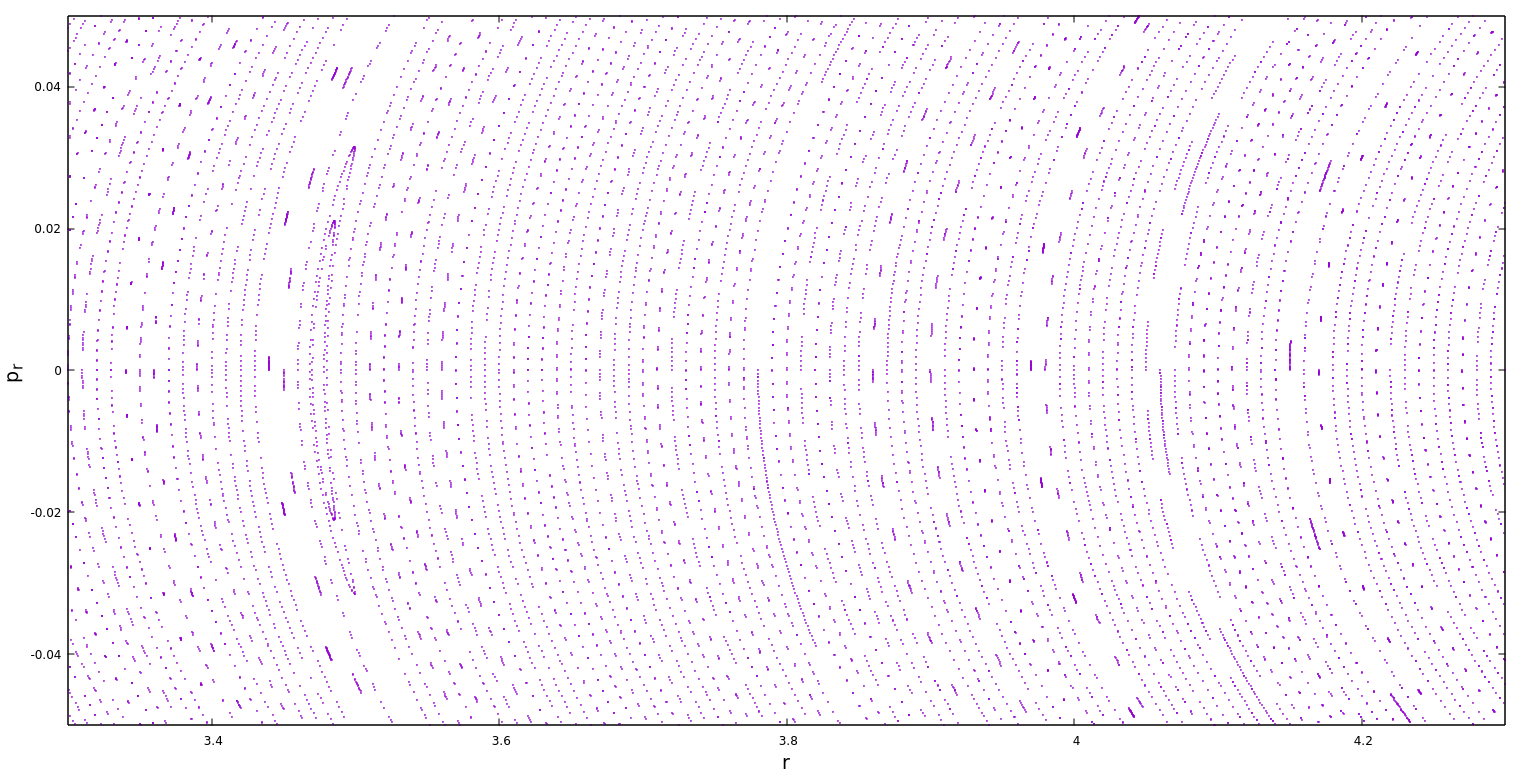}
    \caption{Poincaré section for the parameter set $M = 1.0$, $a = 0.99$, $L_z = 1.2$,  $E = 0.932516$ y $q = 0.3$. It shows the island located at $r = 3.475$.}    
    \label{F:q03a}
\end{figure}

\begin{figure}[H]
    \centering
    \includegraphics[width=0.9\textwidth]{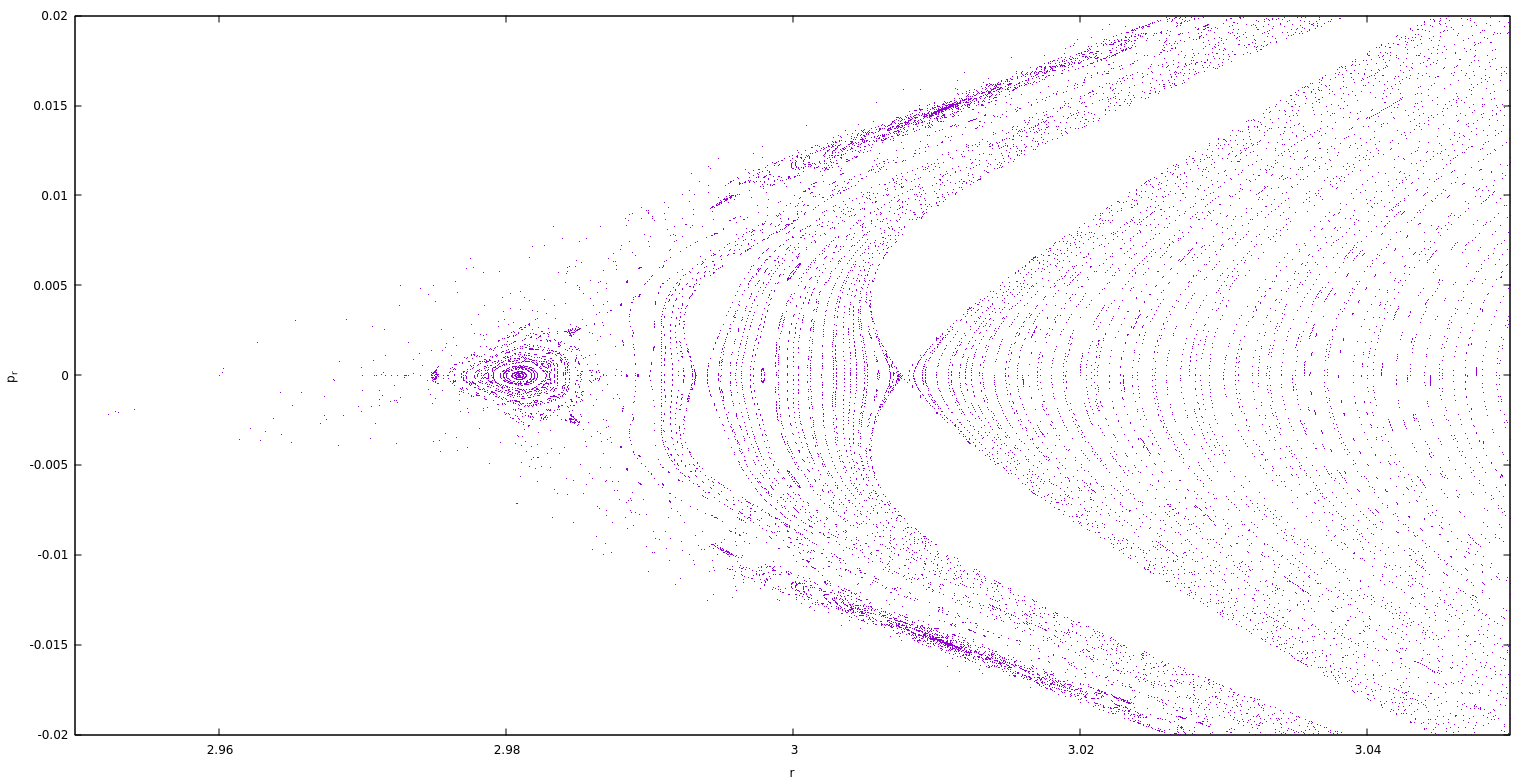}
    \caption{Poincaré section for the parameter set $M = 1.0$, $a = 0.99$, $L_z = 1.2$, $E = 0.932516$ y $q = 0.3$, in the region near the event horizon.}    
    \label{F:q03b}
\end{figure}

\begin{figure}[H]
    \centering
    \includegraphics[width=0.9\textwidth]{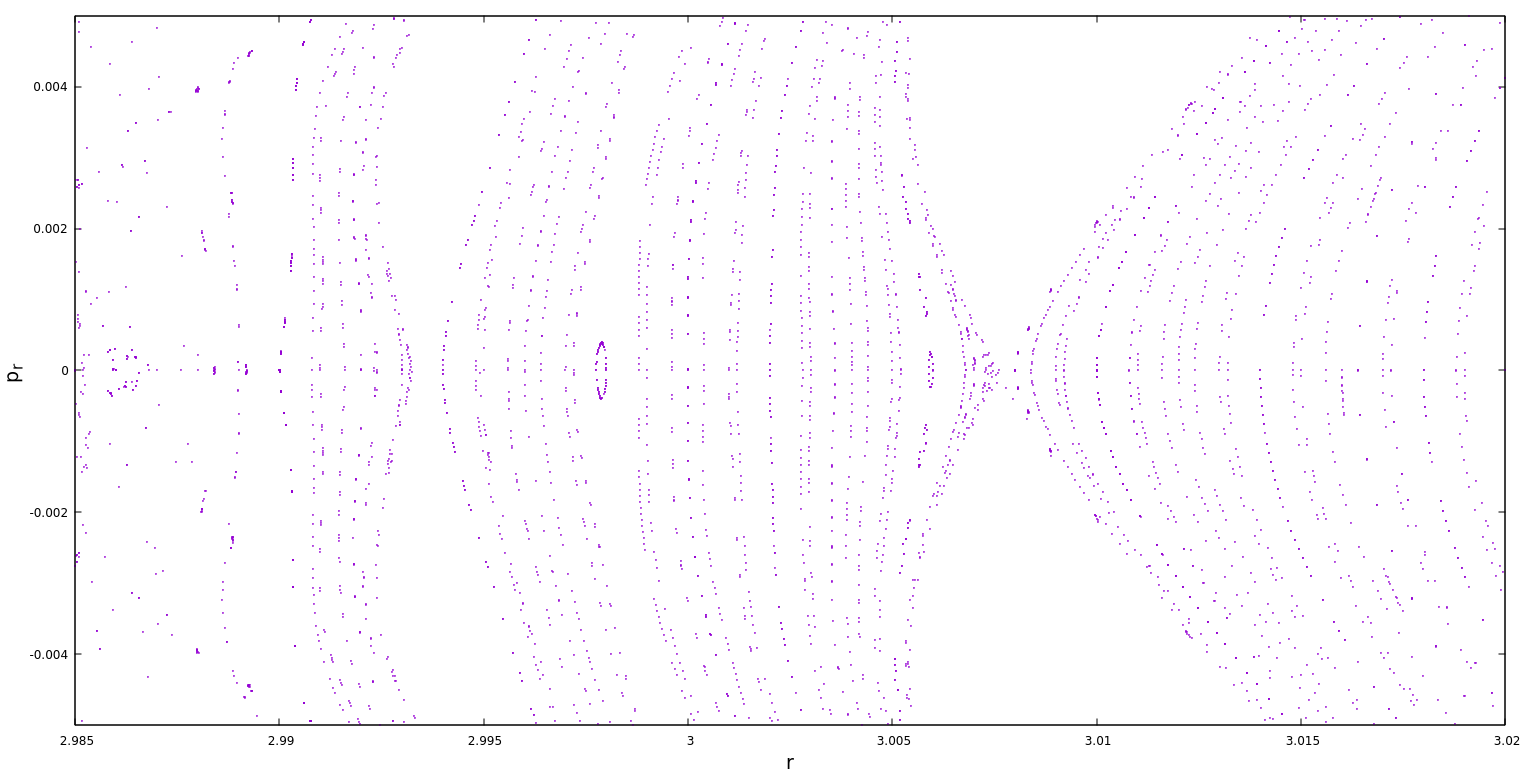}
    \caption{Poincaré section for the parameter set  $M = 1.0$, $a = 0.99$, $L_z = 1.2$,  $E = 0.932516$ y $q = 0.3$. It shows a zoom of the figure \ref{F:q03b} near the interval between hyperbolic points.}    
    \label{F:q03c}
\end{figure}

\begin{figure}[H]
    \centering
    \includegraphics[width=0.9\textwidth]{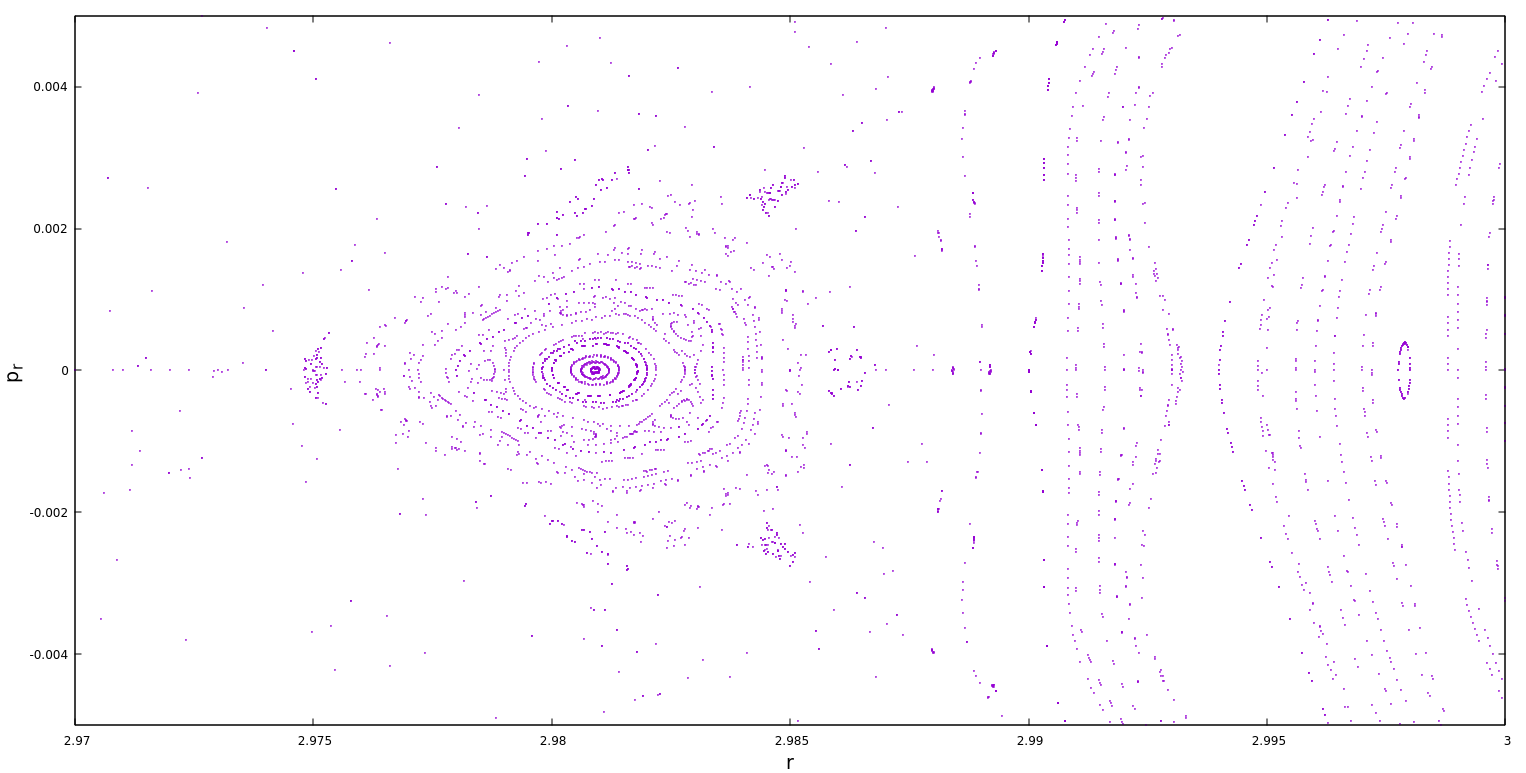}
    \caption{Poincaré section for the near event horizon structure and for parameters  $M = 1.0$, $a = 0.99$, $L_z = 1.2$,  $E = 0.932516$ y $q = 0.3$. It shows a zoom of figure \ref{F:q03b} in the structures near the event horizon.}    
    \label{F:q03d}
\end{figure}

Setting $q = 0.3$ while the rest of the parameters remain constant, new structures arising in the phase space are seen. In the region nearest the event horizon, as previously said, the gravity is extremely strong, and will show a high density of new structures. In the Poincaré sections of figures \ref{F:q03a}, \ref{F:q03b}, \ref{F:q03c} and \ref{F:q03d}, it is easy to verify a new islands and other structures crossing the $p_r = 0$ axis. The large island of figure \ref{F:q03b} is surrounded by higher order island and a scattered points sea. Those scattered points represents the geodesics that have transformed into strange attractors and are chaotic regions. Moving along the $p_r = 0$ to the right, small islands and two hyperbolic points separating regions are identified. Particularly, in figure \ref{F:q03c} we could see that broken tori are actually small chains of Birkhoff islands, and the hyperbolic point at $r = 3.008$ separates the region of the broken tori from the slightly deformed ones.  The large white spaces in those figures are embedded islands orbiting the main one.

In figure \ref{F:q03b}, it is found a new large island that is surrounded by several higher order islands and small chaotic orbiting regions. Those chaotic orbits are unstable, eventually, they will fall into the compact object. However, those orbits will show the stickiness phenomenon, i.e. they will remain attached to the stable orbits for a period of time before separating due to the growing non-linearities. In order to better analyze the chaotic regions, we will use the rotation number along the $r$-axis of the Poincaré section. The interval where the variation is very high is a direct prove of chaos in the geodesic \cite{LukesZpoyVoorhees, AngulardynamicsVoglis, HowtoObserveNonKerr, countopouloslukes2014non, contopoulosgerakopoulos, contopoulosorderandchaos}. The rotation number is a powerful test because the stickiness will not affect the final result. However, it is necessary to make the simulations for a long period of time.

Figure \ref{F:q03a} shows the farthest detected island from the event horizon, located at $p_r = 0$ and $r = 3.475$. This region is absent of satellites or scattered points and therefore the surrounding region in the phase space will be stable. Moreover, the rotation number will be a constant rational number along the extension of the island. However, the existence of such islands in the gravitational field near the event horizon is the first indication that the dynamical system is non-integrable and the emerging of chaos getting close to the event horizon.

Analyzing the orbits for smaller radial distances in figures \ref{F:q03b} and \ref{F:q03c}, several new structures like hyperbolic points are found located at $r = 3.00775$ and $p_r = 0$. Additionally, this particular hyperbolic point is the boundary between two regions with different behaviors. For higher $r$ the orbits are quite stable with few narrow islands. For smaller $r$, the region exhibit several structures like Birkhoff islands, satellites and scattered points. 

Figure \ref{F:q03d} shows the closest structure to the event horizon, centered at $p_r = 0$ and $r = 2.981$. This small island is surrounded by several satellites and a sea of scattered points and therefore surrounded by chaotic orbits. Those chaotic geodesics are strange attractors in the phase space, most of them are unstable and eventually will fall to the event horizon. However, these orbits will remain attached to a stable orbit because of the stickiness.

\subsection{Mass Quadrupole Moment $q = 0.7$}

\begin{figure}[H]
    \centering
    \includegraphics[width=0.9\textwidth]{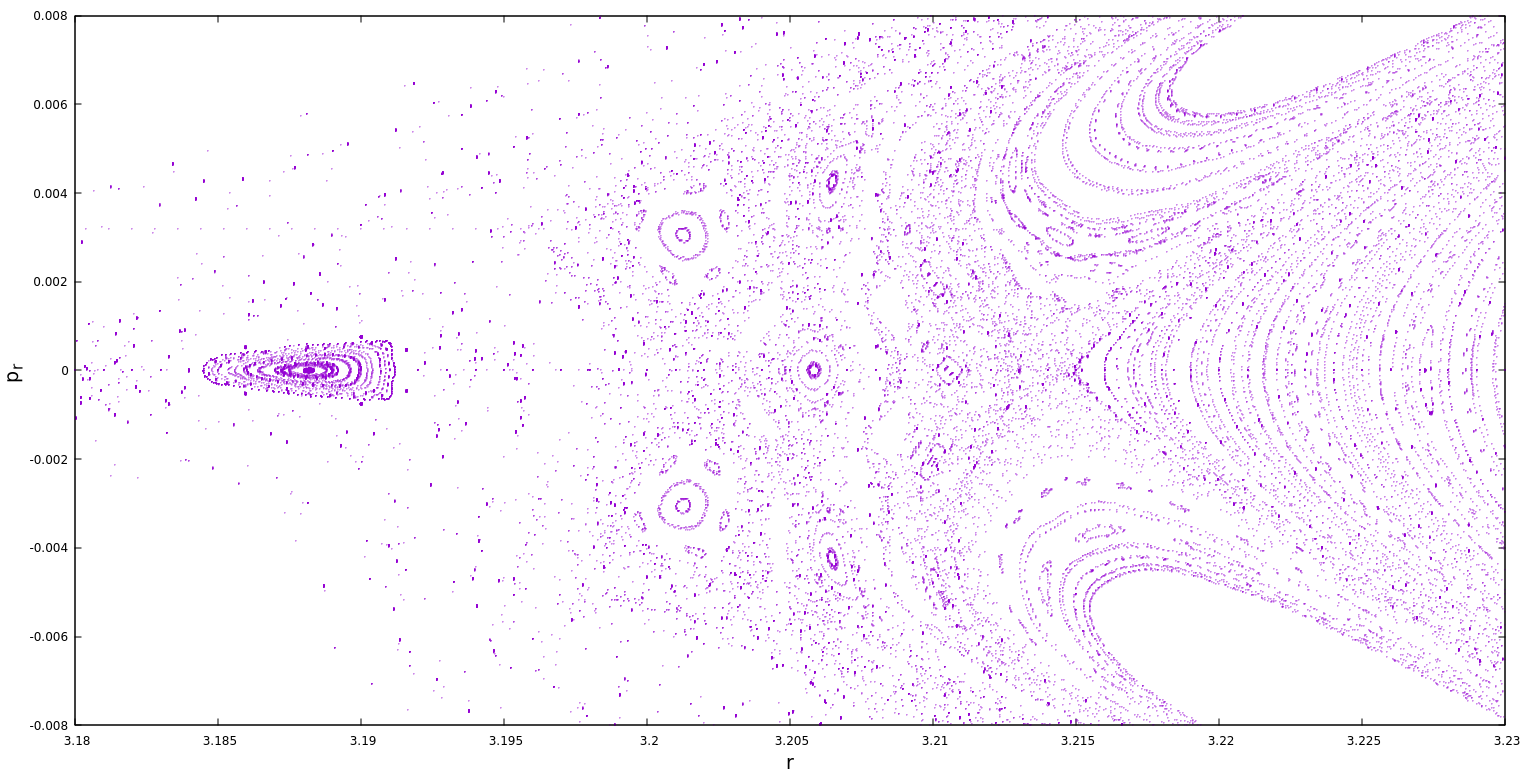}
    \caption{Poincaré section for the event horizon nearest structure and parameters $M = 1.0$, $a = 0.99$, $L_z = 1.2$,  $E = 0.932516$ and $q = 0.7$.}    
    \label{F:q07}
\end{figure}

Figure \ref{F:q07} shows the most relevant structures in the phase space if we set $q = 0.7$. Due to the torus breaking as the $q$ parameter is increased, the center of the main island of stability will move to $r = 6.144$. The same goes for other structures like the island closest to the event horizon now located at $r = 3.1882$. There are several resonances outside the axis $p_r$ surrounded by satellites, for example $(r, p_r)$ = $(3.2012, \, \pm 0.00303)$, $(r, p_r) = (3.206, \, 0)$. In section \ref{Sec:Rot} these resonances are identified using the rotation number. At $(r, p_r) = (3.2106, \, 0)$ there is a deformed island together with a chain of tiny islands. Additionally, the hyperbolic point located at $r = 3.215$ is actually broken; it is composed by several tiny islands. This is a direct consequence of the effect of a highly deformed gravitational source over the spacetime surrounding it, and producing chaotic geodesics. Finally, the \emph{lobes} that symmetrically appear to both sides of the $p_r$ axis and separated by the \emph{hyperbole} centered at $r = 3.215$ are surrounded by higher order island and a chaotic region.

\subsection{Characteristic of $u_0$ and $u_0 '$.}

The characteristic is the position of a periodic orbit as a function of a single parameter of the system while keeping fixed the remaining parameters \cite{contopoulosgerakopoulos}. Figure ref{F:ex1} shows the position of the center of the main island of stability $u_0$.  Table \ref{T:Carac.elipt} shows the decrease of the characteristic of this elliptical point as the mass quadrupole moment increases.

\begin{figure}[H]
    \centering
    \includegraphics[width=0.9\textwidth]{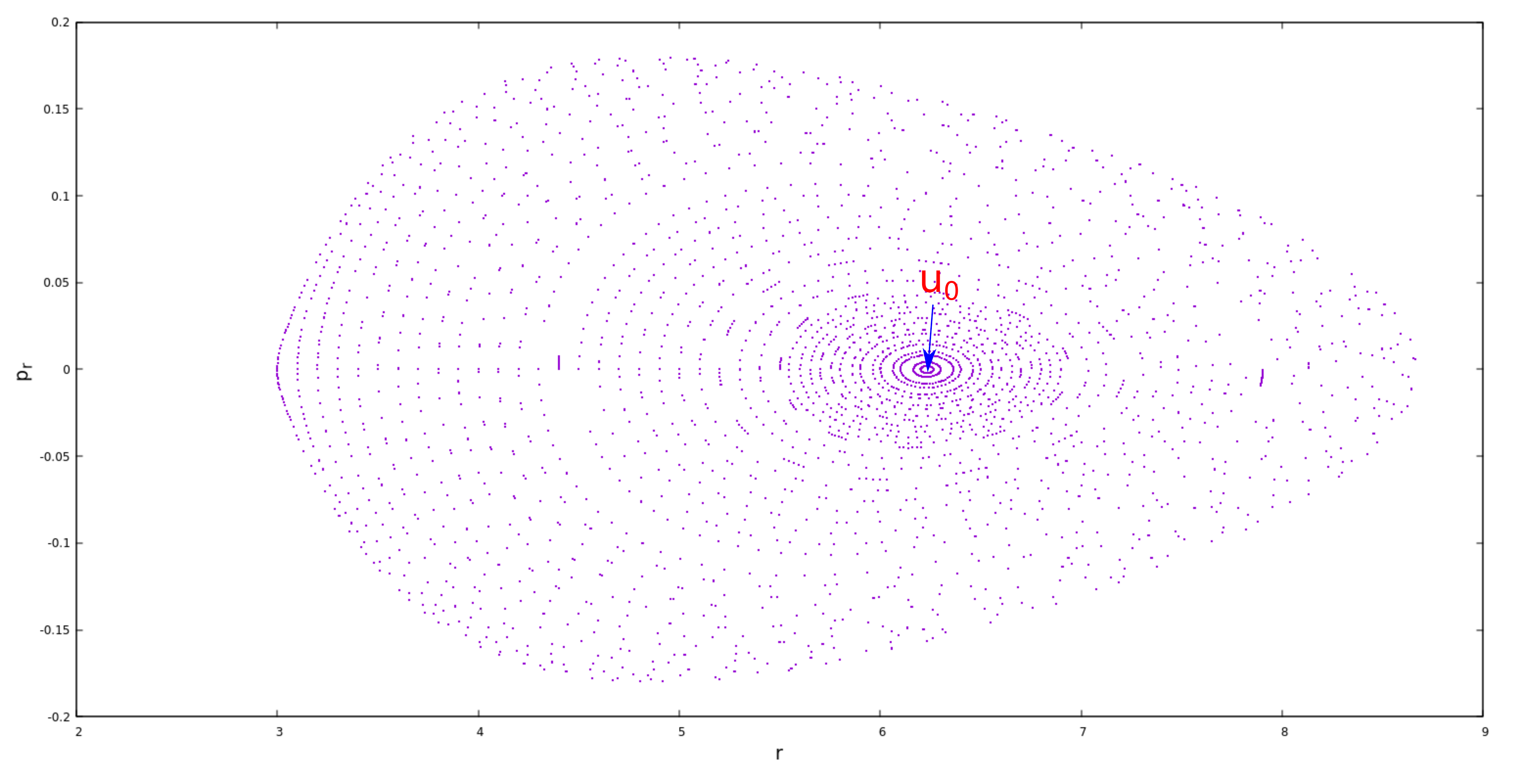}
    \caption{Position of the center of the main stability island $u_0$ for $q = 0.3$.}    
    \label{F:ex1}
\end{figure}

\begin{table}[H]
\caption{Characteristic of the center of main island of stability $u_0$ as varying the mass quadrupole moment.}
\label{T:Carac.elipt}
\begin{center}
\begin{tabular}{ c  c  c  }
\hline
\textbf{Mass Quadrupole} & \textbf{Position of } \\
\textbf{ Moment $q$}     & \textbf{ $u_0$ }    &   \\
\hline

    $0$       &   $6.294$  \\
    $0.01$    &   $6.293$  \\
    $0.1$     &   $6.274$  \\
    $0.2$     &   $6.253$  \\
    $0.3$     &   $6.232$  \\
    $0.4$     &   $6.211$  \\
    $0.5$     &   $6.189$  \\
    $0.6$     &   $6.167$  \\
    $0.7$     &   $6.144$  \\
    $0.8$     &   $6.121$  \\
    $0.9$     &   $6.098$  \\
    $0.95$    &   $6.086$  \\
\hline
\end{tabular}
\end{center}
\end{table}

\begin{figure}[H]
    \centering
    \includegraphics[width=0.9\textwidth]{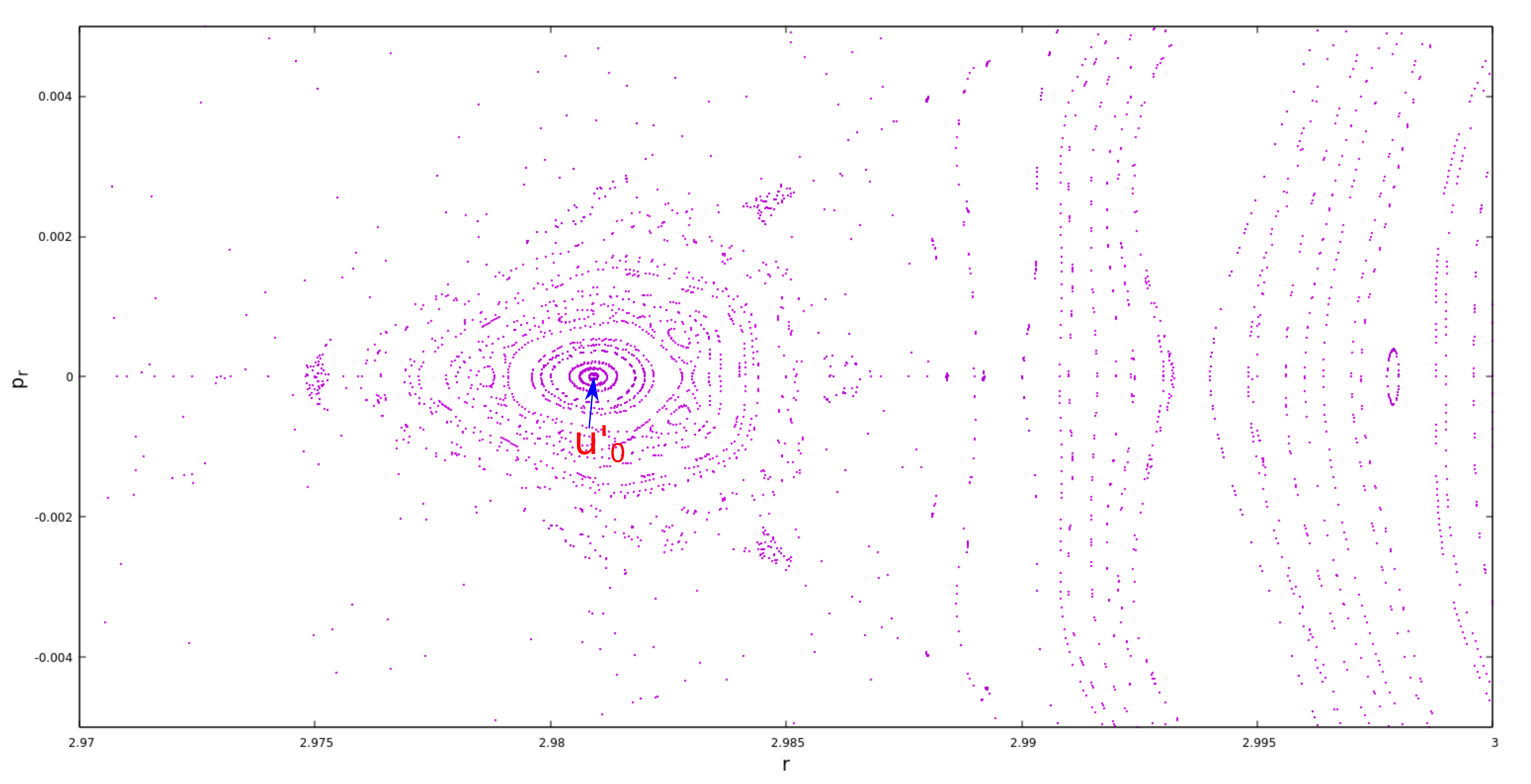}
    \caption{Position of the center of the nearest event horizon island $u'_0$ for $q = 0.3$.}    
    \label{F:ex2}
\end{figure}

The center of the island closest to the event horizon $u_0 '$ is seen in figure \ref{F:ex2}. Contrasting with the characteristic of $u_0$, the position of $u_0 '$ increases, as seen in table \ref{T:estruct}.

\begin{table}[H]
\caption{Characteristic of the center of the nearest event horizon island $u'_0$.}
\label{T:estruct}
\begin{center}
\begin{tabular}{ c  c    }
\hline
\textbf{Mass Quadrupole} & \textbf{Position of } \\
\textbf{ Moment $q$}     & \textbf{ $u'_0$}   \\
\hline
    $0$       &   -      \\
    $0.01$    &   $2.7820$  \\
    $0.1$     &   $2.8619$ \\
    $0.2$     &   $2.9228$ \\
    $0.3$     &   $2.9809$ \\
    $0.4$     &   $3.0362$ \\
    $0.5$     &   $3.0890$ \\
    $0.6$     &   $3.1396$ \\
    $0.7$     &   $3.1882$ \\
    $0.8$     &   $3.2350$ \\
    $0.9$     &   $3.2803$ \\
    $0.95$    &   $3.3025$ \\
\hline
\end{tabular}
\end{center}
\end{table}

Since the mass quadrupole moment $q$ represents the deviation parameter from the Kerr metric, the evolution of the new structures in the phase space is directly related to this parameter. Many of the nearest compact object tori will fragment as $q$ increases, they transform into small chains of islands instead been completely destroyed. It is important to note that to construct the rotation number it is necessary to find the characteristic of $u_0$ as shown in the next section.

\section{Rotation Number} \label{Sec:Rot}

\begin{figure}[H]
    \centering
    \includegraphics[width=0.9\textwidth]{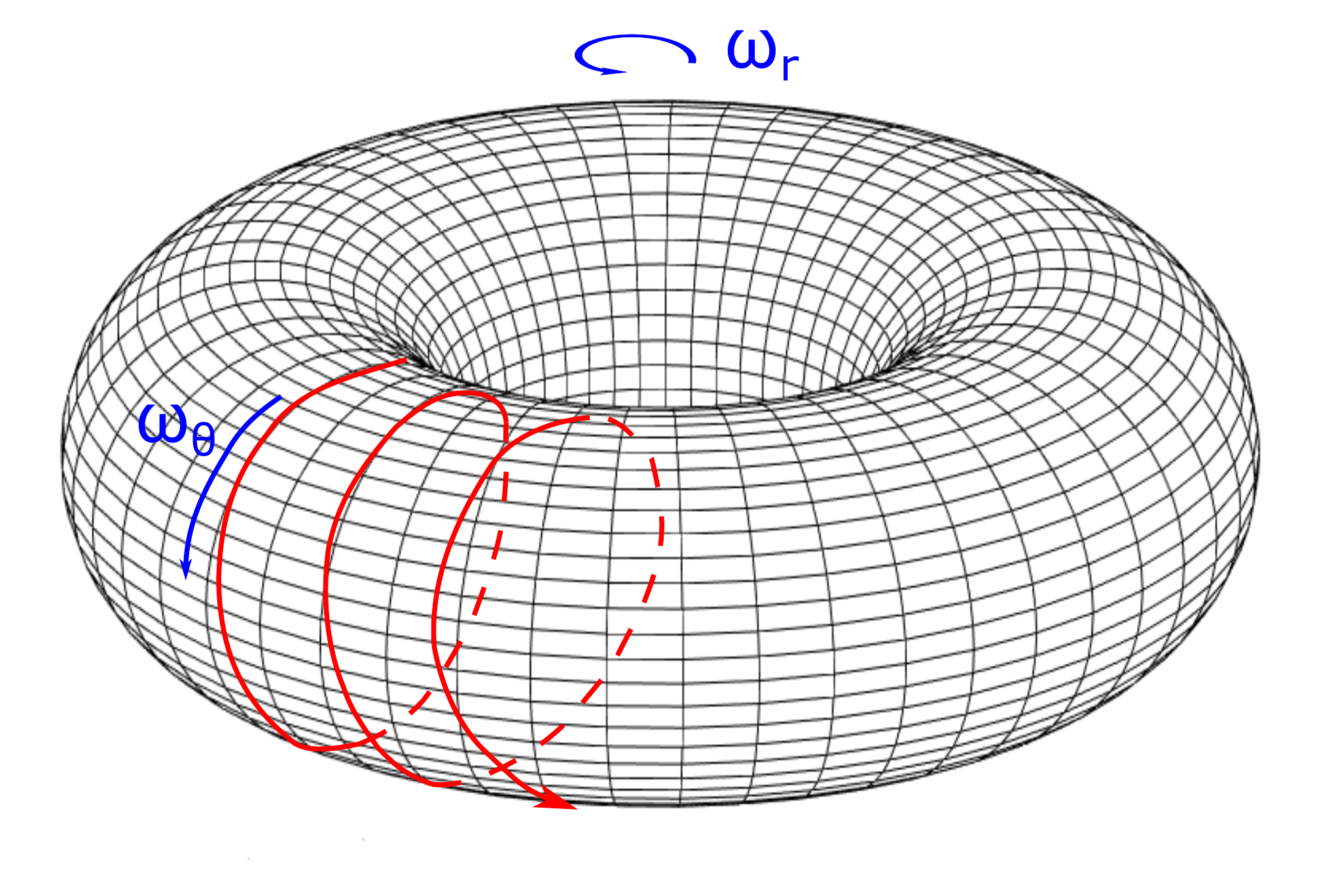}
    \caption{Motion of a particle bounded to a torus surface. The frequencies $\omega_r$ and $\omega_\theta$ are used to calculate the rotation number.} \label{F:toro}
\end{figure}

The rotation number $\nu_\theta$ is defined as the ratio $\omega_r / \omega_\theta$ of frequencies, as shown in the figure \ref{F:toro}. It is called a resonance when $\nu_\theta$   is a rational number and this corresponds to an island in the Poincaré section. Moreover, if the rotation number is taken as a function of the initial radial distance $r_0$, then $\nu_\theta$ will be constant along the extension of the island. On the other hand, if there are hyperbolic points crossing the region where the  $\nu_\theta$ is calculated, then the rotation number will have a jump, or discontinuity. Finally, when $\nu_\theta$ varies rapidly during some interval, then that region corresponds to a chaotic one. The chaotic regions actually have a fractal structure with tiny constant intervals of $\nu_\theta$ submerged into a chaotic region \cite{contopoulosgerakopoulos,contopoulosorderandchaos,countopouloslukes2014non,HowtoObserveNonKerr,LukesZpoyVoorhees}.

The rotation number is a powerful test to classify a region in the Poincaré section \cite{contopoulosgerakopoulos,contopoulosorderandchaos,countopouloslukes2014non,HowtoObserveNonKerr,LukesZpoyVoorhees,MoisesSantos}. Easily, it could identify stability islands, hyperbolic points or scattered points along some axis. Particularly, the regions where the chaotic orbits remain stuck to some stable geodesic are easy identifiable \cite{MoisesSantos}. 

\begin{figure}[H]
    \centering
    \includegraphics[width=0.7\textwidth]{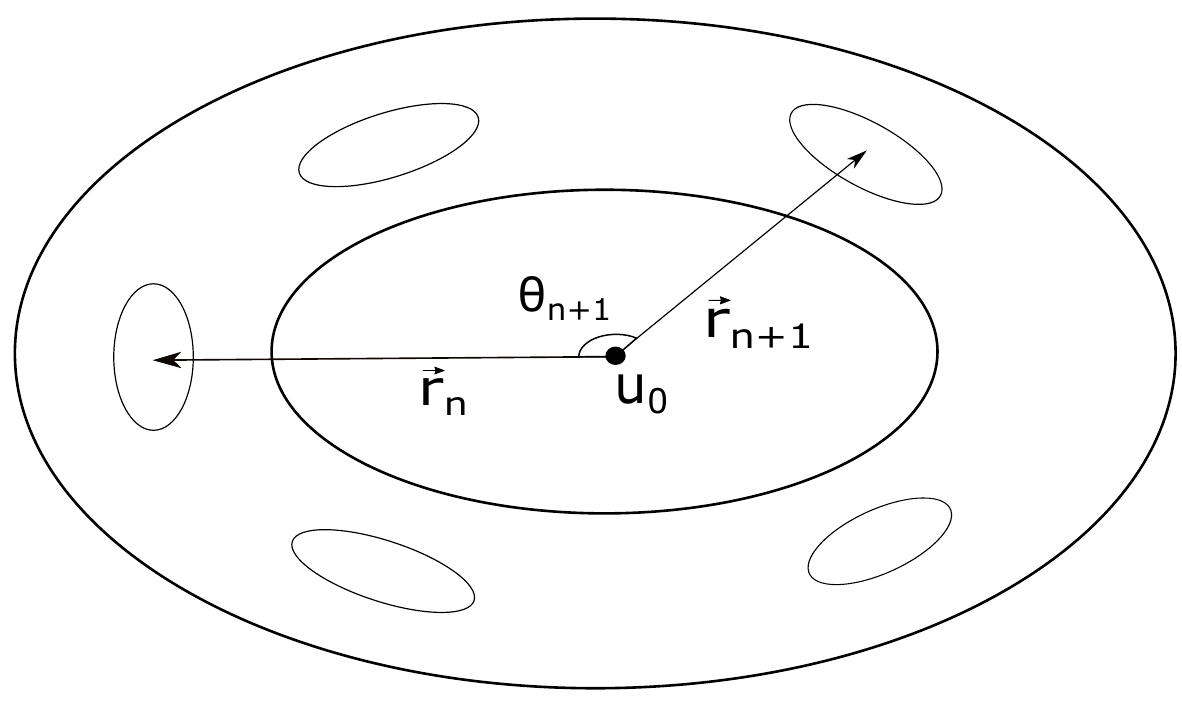}
    \caption{Angle $\theta_{n+1}$ formed between the position of vectors $\vec{r}_n$ and $\vec{r}_{n+1}$ of the center of islands. The $u_0$ point represents the center of the main island of stability. This example illustrates a resonance $2/5$.}    
    \label{F:rot0}
\end{figure}

However, to calculate the rotation number by definition is difficult, but there is a simple algorithm used to calculate the rotation number by taking the points in the Poincaré section. Firstly, it is necessary to identify the center of the main island of stability, as we have done in table \ref{T:Carac.elipt}. Secondly, we calculate the angle $\theta_{n+1} = \measuredangle(\vec{r}_{n+1} , \vec{r}_n) $ between two consecutive points of the Poincaré section for some fixed initial value $r_0$, as shown in figure (\ref{F:rot0}). Although this figure shows only the centers, it could be calculated for some other consecutive points of the Poincaré section. The position of the Poincaré section points $\vec{r}_n$ are calculated using $u_0$ as the reference. Then the rotation number is

\begin{equation}
    \label{E:rot}
    \nu_\theta = \lim_{n \to \infty} \frac{1}{2\pi n} \sum _{j = 1} ^{n} \theta_{j}
\end{equation}

If we construct the rotation number as a function of $r_0$ along $p_r = 0$ axis then chaotic regions could be detected. The computational algorithm for $\nu_\theta(r)$ musts take into account repeated values for the same island so it musts not over count the same point.

This algorithm is general in the sense that if we change the metric then it is not necessary to change the algorithm. There are other ways to calculate the rotation number; nevertheless, they are dependent on the dynamical system equations, for example the Fourier analysis. These ways to calculate the rotation number are explored in \cite{AngulardynamicsVoglis, FrequencyFourier, FrequencyFourierLaskar}

\begin{figure}[H]
    \centering
    \includegraphics[width=0.9\textwidth]{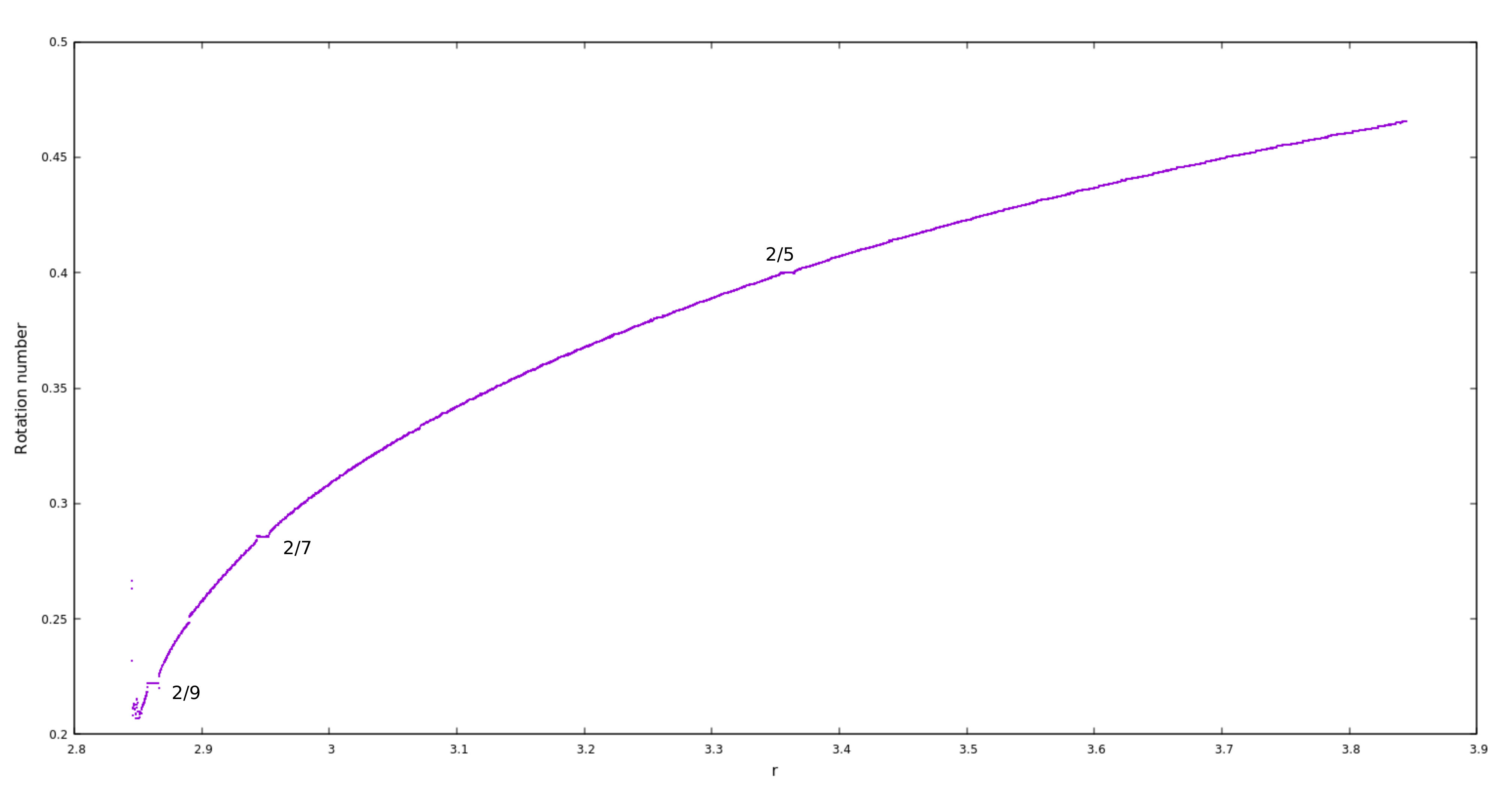}
    \caption{Variation for the rotation number along $p_r = 0$ axis for $q = 0.1$, $M = 1.0$, $a = 0.99$, $L_z = 1.2$, $E = 0.932516$.}    
    \label{F:rotq01}
\end{figure}

\begin{figure}[H]
    \centering
    \includegraphics[width=0.9\textwidth]{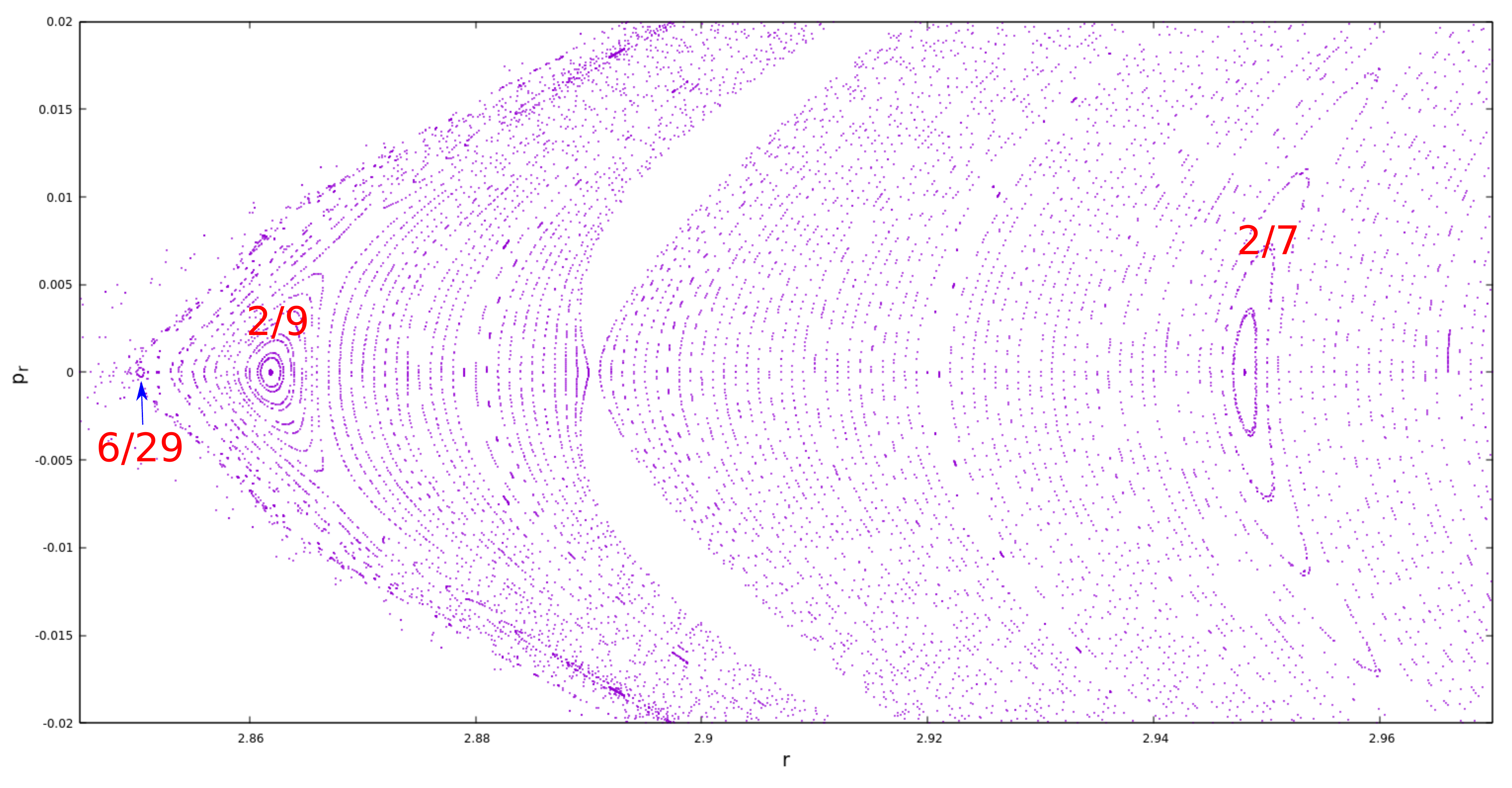}
    \caption{Poincaré section for $M = 1.0$, $a = 0.99$, $E = 0.932516$, $L_z = 1.2$ y $q = 0.1$, showing the respective resonances crossing $p_r = 0$. There are several scattered points surrounding the $6/29$ island.} 
    \label{F:poinq01}
\end{figure}

Figure \ref{F:rotq01} shows the rotation number as a function of the initial radial distance for $\theta = \pi/2$, $p_r = 0$ and $q = 0.1$. There are three obvious resonances $2/9, \, 2/7$ and $2/5$, some jumps and a highly variable interval corresponding to islands, hyperbolic points and a chaotic regions respectively. Figure \ref{F:poinq01} shows the resonances $2/9, \, 2/7$ and, a another less obvious, $6/29$ in the neighborhood of the chaotic region. 

\begin{figure}[H]
    \centering
    \includegraphics[width=0.9\textwidth]{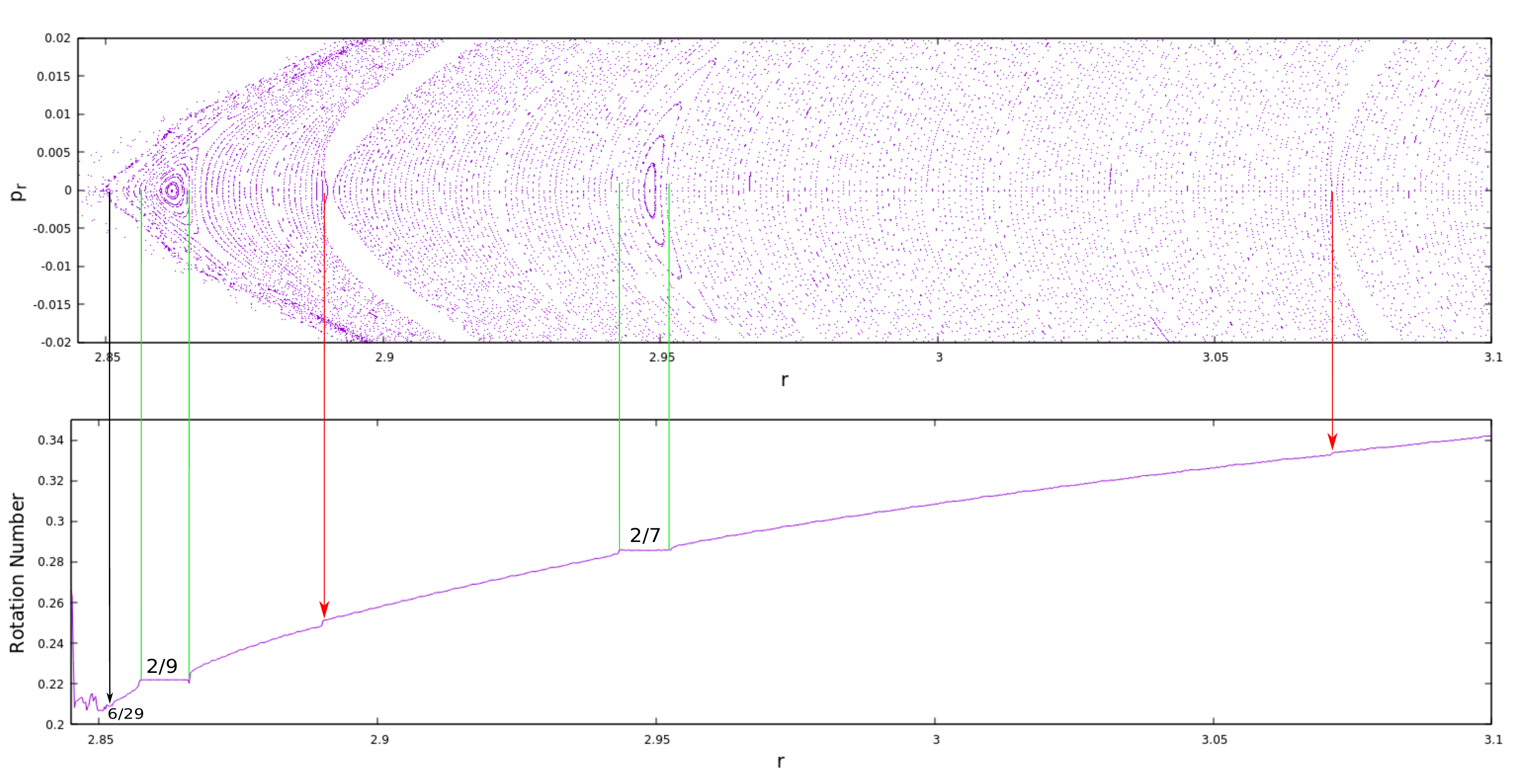}
    \caption{Rotation number along the $p_r = 0$ axis for $q = 0.1$ and the corresponding Poincaré section for the same conditions of Figure \ref{F:poinq01}. The islands width is delimited by green lines, the hyperbolic points are marked by red arrows and the small $6/29$ island by a black arrow.} 
    \label{F:rotq01b}
\end{figure}

Figure \ref{F:rotq01b} shows the Poincaré section with the corresponding rotation number for the same conditions of shown in figure \ref{F:poinq01}. The position of the $6/29$ resonance ($r = 2.8503$) is on the border of the chaotic region and very close to the last stable orbit. For this set of parameters, it is concluded that there is only incipient chaos surrounding the resonance $6/29$, because the rest of the rotation number function is quite monotone, with only other resonances and hyperbolic points present. However, if we increase the deformation of the gravitational source while keeping the rest of the parameters fixed, it is expected new chaotic regions surrounding the islands or in the neighborhood of hyperbolic points.

\begin{figure}[H]
    \centering
    \includegraphics[width=0.9\textwidth]{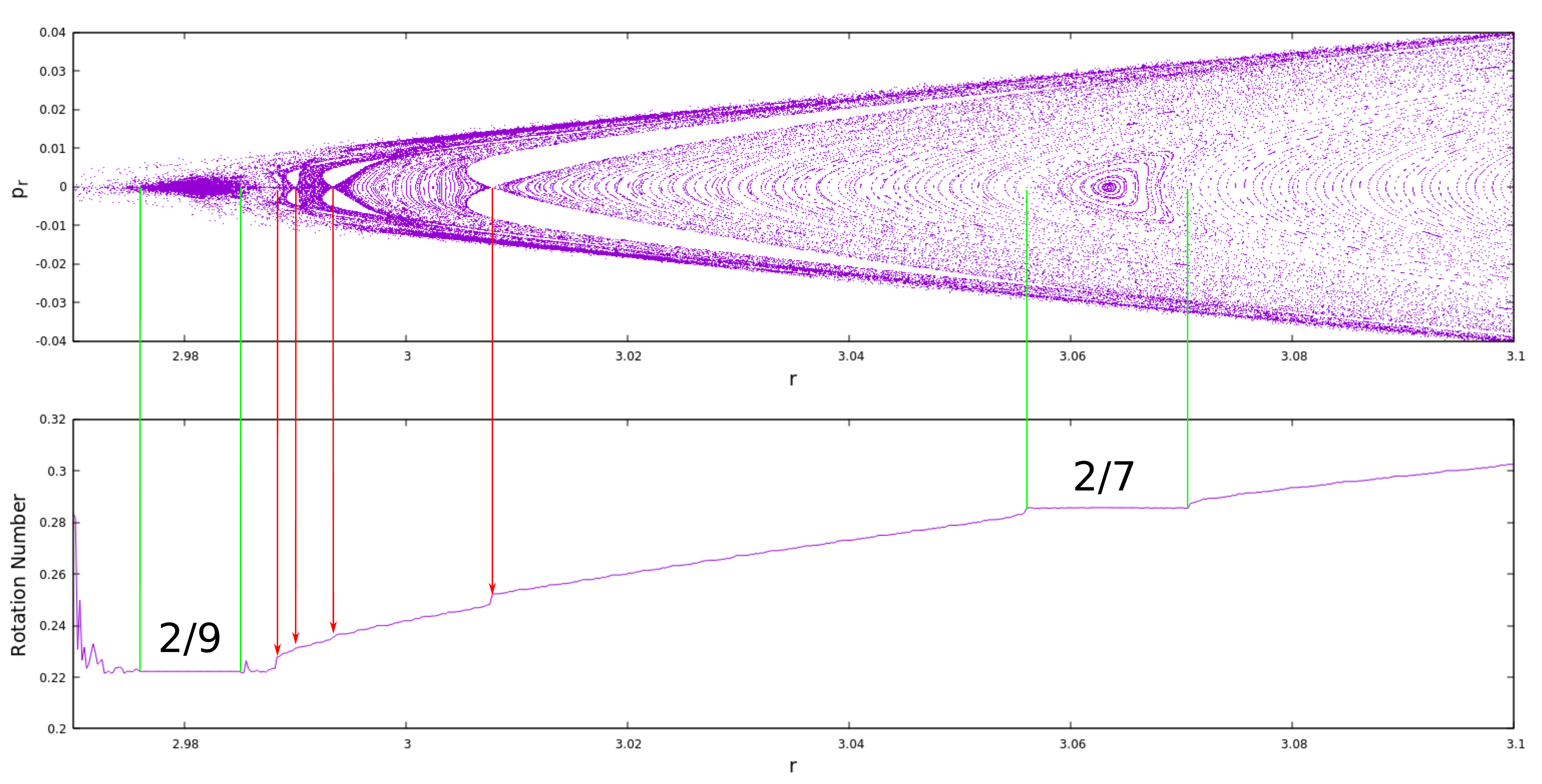}
    \caption{Rotation number along $p_r = 0$ axis for $q = 0.3$ and its corresponding Poincaré section for the same conditions of figure \ref{F:poinq01}. The red arrows show the position of the hyperbolic points with their corresponding jump in the rotation number. The green lines show the widths of the $2/9$ and $2/7$ islands.}    
    \label{F:rotq03b}
\end{figure}

Increasing the mass quadruple moment to $q = 0.3$, we find that the resonances $2/9$ and $2/7$ survive while $6/29$ not. The $2/5$ resonance actually survives too, but it is outside the range of figure \ref{F:rotq03b}. As the tori of the phase space of the resonance $6/29$ breaks out new small structures crossing the $p_r = 0$ axis appear.

\begin{figure}[H]
    \centering
    \includegraphics[width=0.9\textwidth]{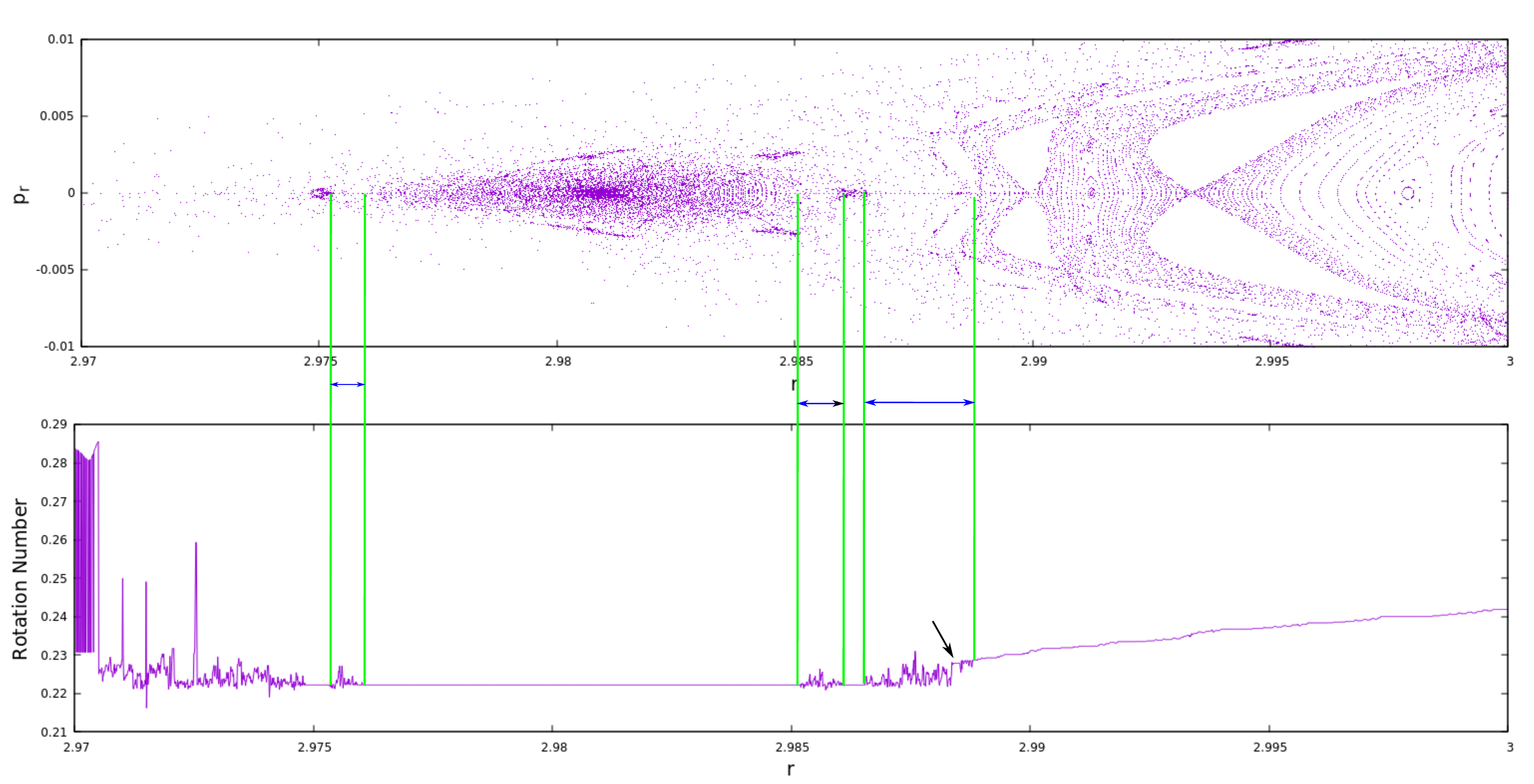}
    \caption{Zoom of figure \ref{F:rotq03b} showing the rotation number along $p_r = 0$ for $q = 0.3$ with its corresponding Poincaré section for the same conditions of figure \ref{F:poinq01}. The green lines with double blue arrow show three chaotic regions. Additionally, there is another chaotic region on the left side of the figure which extends to the event horizon. The black arrow points to a narrow resonance submerged in the chaotic region.}    
    \label{F:rotq03d}
\end{figure}

A detail of the rotation number is presented in figure \ref{F:rotq03d}, four chaotic zones are identified. The first one extends from the event horizon until the border of the island centered at $r = 2.975$. It is characterized from being the largest region, and it have the greatest variations of the rotation number. The next chaotic region is also adjacent to the island centered at $r = 2.975$ and extends to the island centered at $r = 2.9809$. This chaotic region is quite narrow compared with the others. The third chaotic region begins in $r = 2.985$, just next to the $2/9$ resonance and finishes with the border of the satellite of $2/9$. The last chaotic region is larger than the second and third regions, and extends from the previous satellite until a small island very near to the hyperbolic point of $r = 2.99$. Additionally, figure \ref{F:rotq03d} reveals other resonances crossing $p_r = 0$. The alternation between chaotic and stable regions is characteristic of chaotic dynamical systems, the small stable regions as surrounded by a sea of chaos. Finally, the fourth chaotic region has small resonances submerged, but they are small and nearly unidentifiable from the chaotic background. As an additional remark, the resolution of the figure \ref{F:rotq03d} is of the order of $1.0 \times 10^{-5}$.

\begin{figure}[H]
    \centering
    \includegraphics[width=0.9\textwidth]{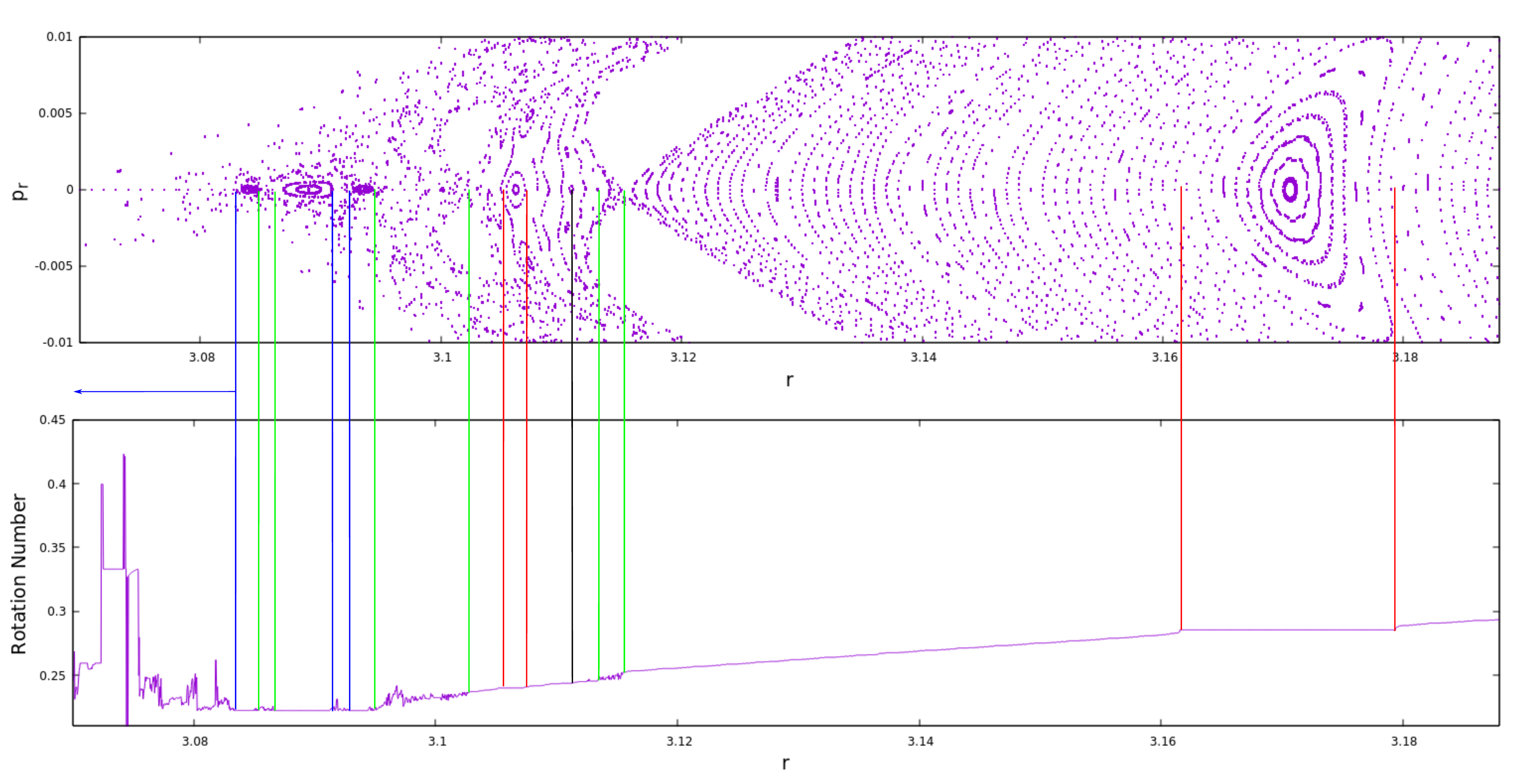}
    \caption{Rotation number along the $p_r = 0$ axis for $q = 0.5$ with the corresponding Poincaré section for the same conditions of figure \ref{F:poinq01}. The pairs of blue lines and green lines show the chaotic regions and the red lines show the extension of two resonances crossing $p_r = 0$. The black line points at the center of a small resonance.}    
    \label{F:rotq05}
\end{figure}

Figure \ref{F:rotq05} presents the behavior of the rotation number over the $p_r = 0$  axis and for $q = 0.5$. This figure shows new structures at the left of the hyperbolic point of $r = 3.116$. At this very same point the chaotic regions ends, for radial distance greater than that, the behavior is regular with only few islands. The resonance at the left of the hyperbolic point and marked by a pair of red lines corresponds to $6/25$ and it is not present in the Poincaré section of $q = 0.3$. The other island marked by two red lines at the right of the hyperbolic point is the $2/7$. Finally, the pair of blue lines and green lines shows the different chaotic regions. 

\begin{figure}[H]
    \centering
    \includegraphics[width=0.9\textwidth]{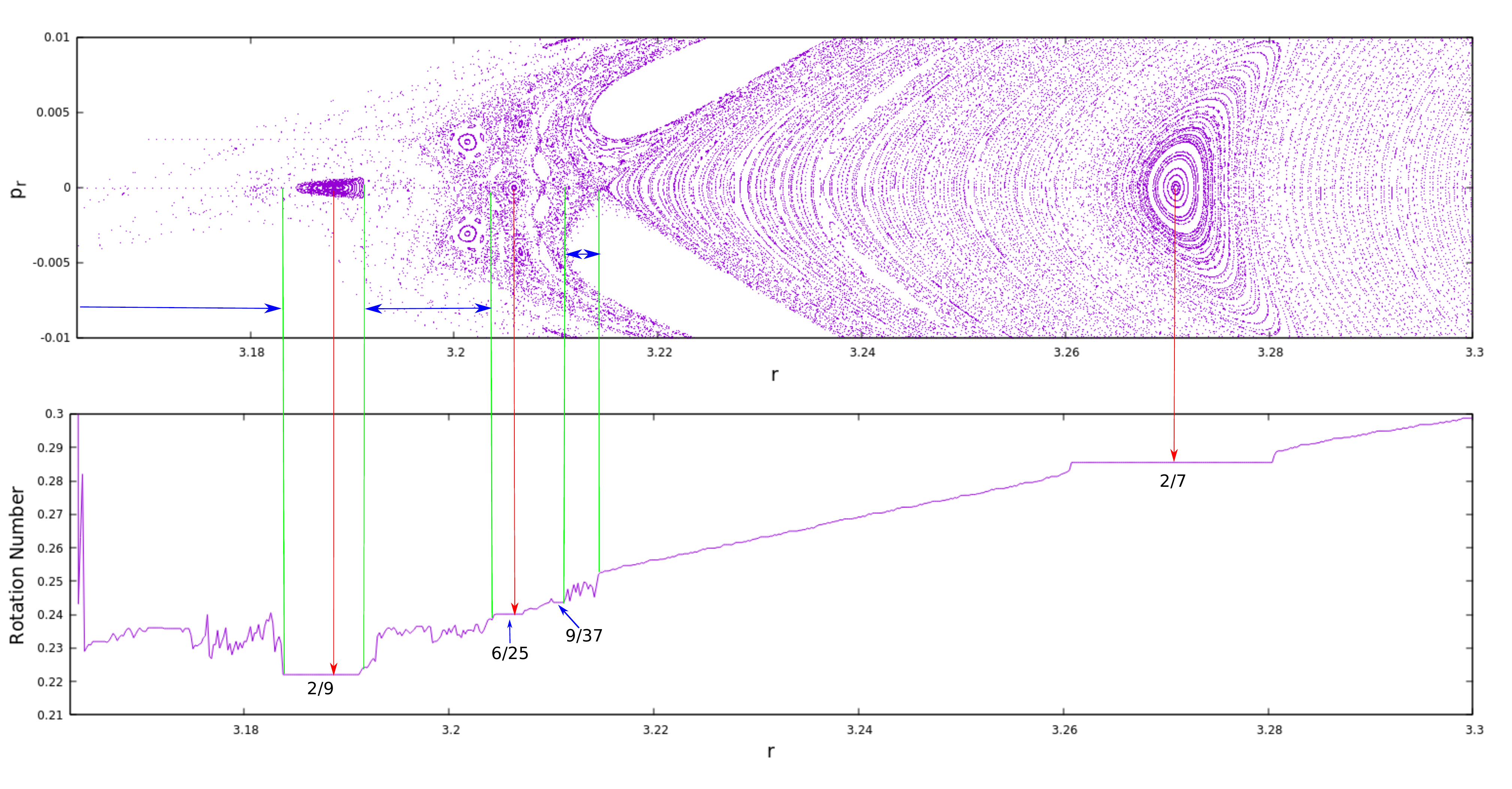}
    \caption{Rotation number along the line $p_r = 0$ for $q = 0.7$ with the corresponding Poincaré section for the same conditions of figure \ref{F:poinq01}. The green lines show chaotic intervals and red arrows show the center of the $2/9, \, 6/25, \, 9/37$ and $2/7$ resonances.}    
    \label{F:rotq07}
\end{figure}

Figure \ref{F:rotq07} shows the rotation number for $q = 0.7$. There are four chaotic regions, the one that extends from the event horizon to the border of the $2/9$ island, one that extends from the $2/9$ to the $6/25$ resonances, another from $6/25$ to $9/37$, and the last one from the border of $9/37$ to the hyperbolic point. Unlike previous cases, in this one there is only one monotonically increasing region between islands $6/25$ and $9/37$. 

\begin{figure}[H]
    \centering
    \includegraphics[width=0.9\textwidth]{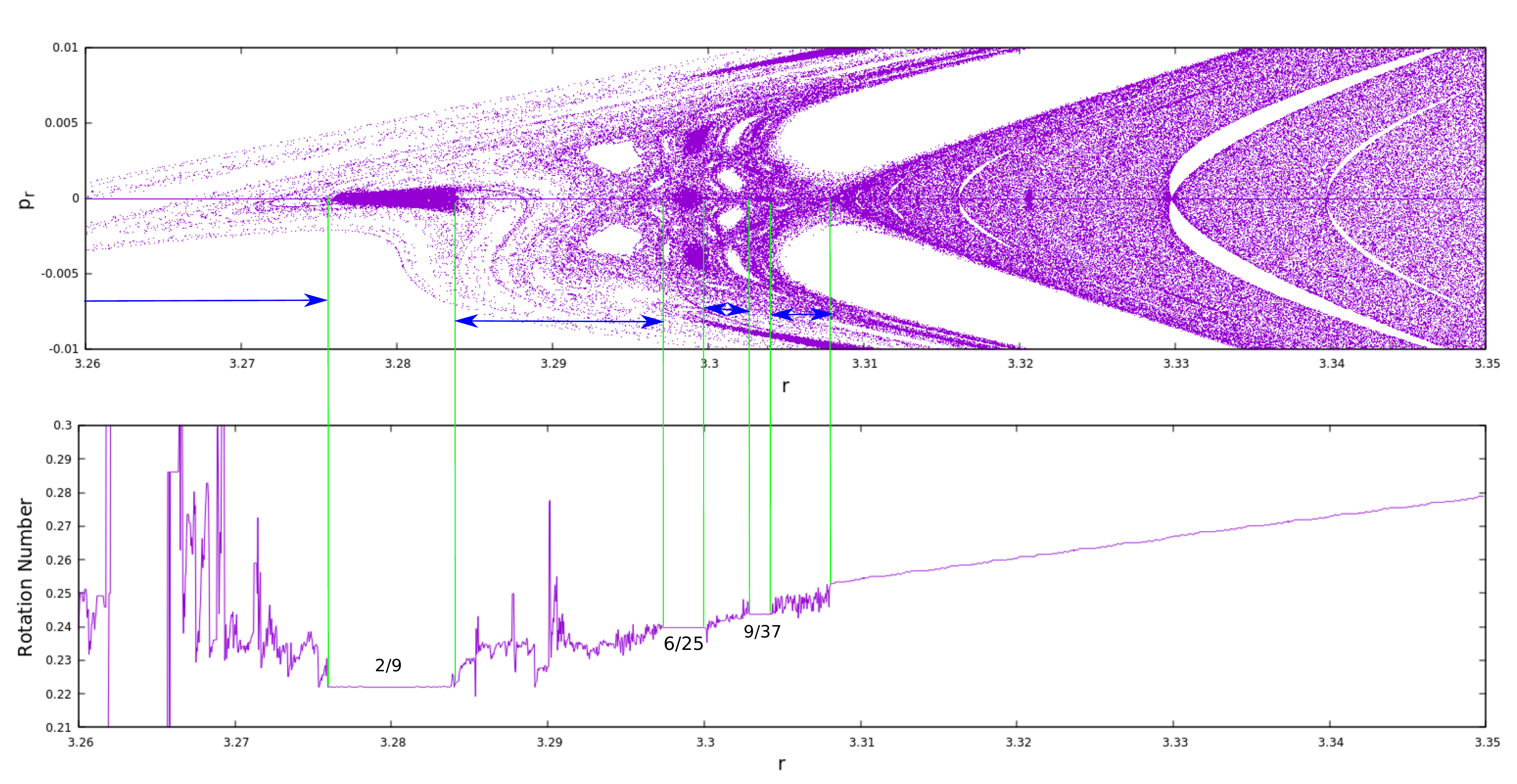}
    \caption{Rotation number over the line $p_r = 0$ for $q = 0.9$ with the corresponding Poincaré section for the same conditions of figure \ref{F:poinq01}. The green lines show the chaotic zones. Additionally, the $2/9, \, 6/25$ and $9/37$ resonances crossing $p_r = 0$ are shown.}    
    \label{F:rotq09}
\end{figure}

\begin{figure}[H]
    \centering
    \includegraphics[width=0.9\textwidth]{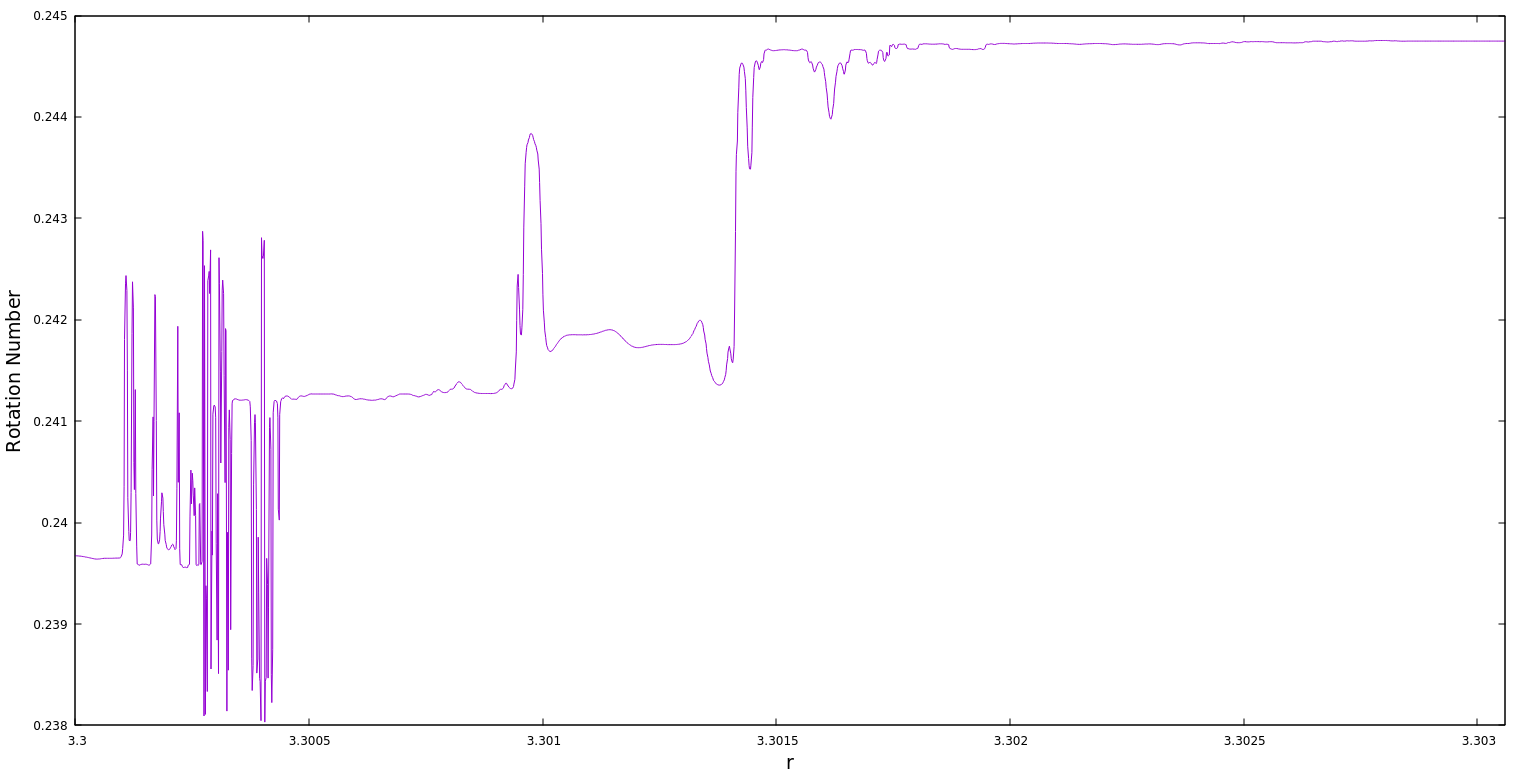}
    \caption{Detail of figure \ref{F:rotq09} in the chaotic interval located between the $6/25$ and $9/37$ resonances.}    
    \label{F:rotq09islas}
\end{figure}

Figure \ref{F:rotq09} shows the evolution of the rotation number and the structures in the Poncaré section for $q = 0.9$. The resonances $2/9, \, 6/25$ and $9/37$ still exists; however, those resonances are stable regions fully surrounded by chaos. Figure \ref{F:rotq09islas} shows the chaotic region between $6/25$ and $9/37$ resonances. There are several small resonances surrounded by rapidly oscillating intervals of the rotation number.

\begin{figure}[H]
    \centering
    \includegraphics[width=0.9\textwidth]{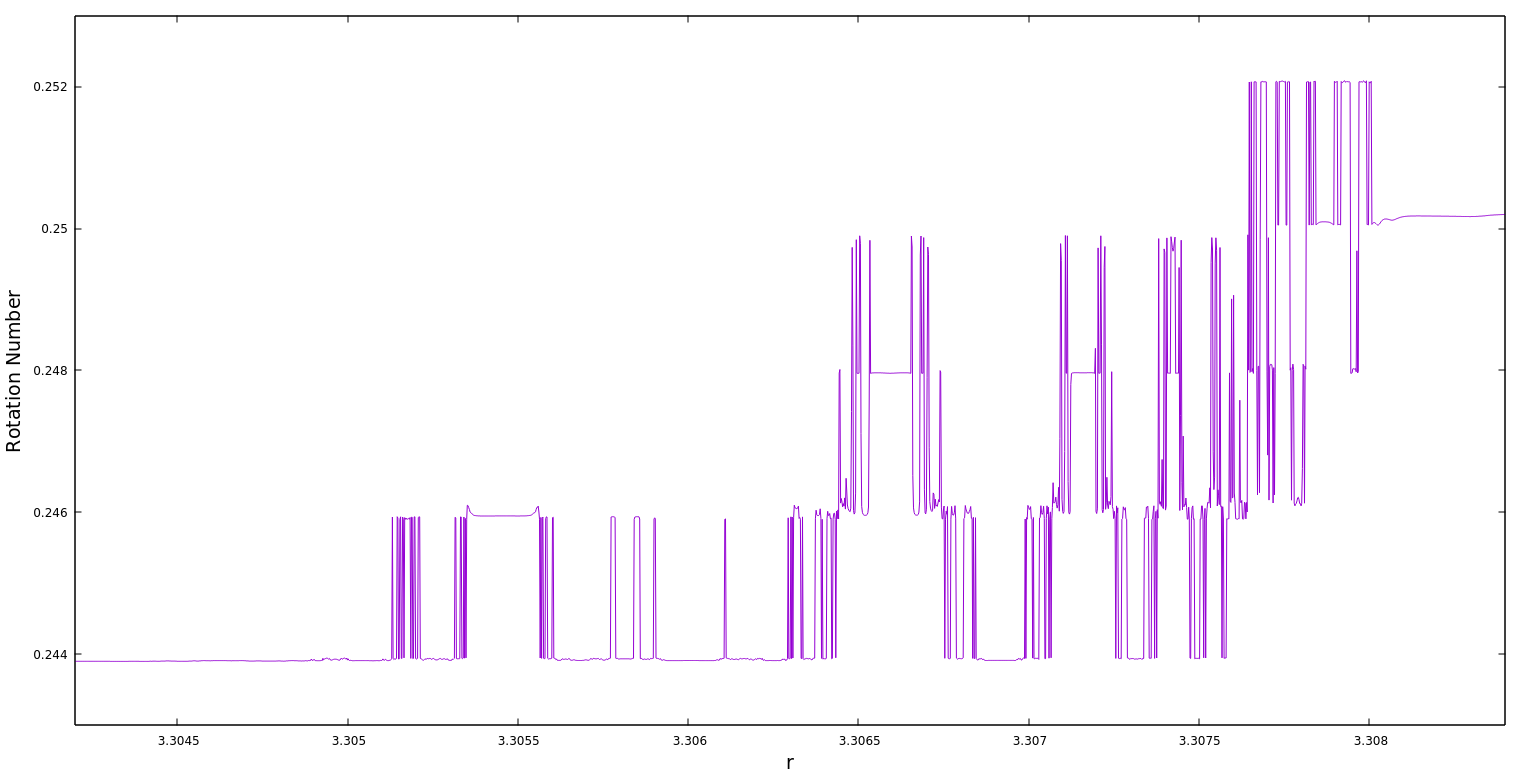}
    \caption{Detail of figure \ref{F:rotq09}. Rotation number in the chaotic interval located between the $9/37$ resonance and the hyperbolic point at $r = 3.30812$.}    
    \label{F:rotq09a}
    \end{figure}

\begin{figure}[H]
    \centering
    \includegraphics[width=0.9\textwidth]{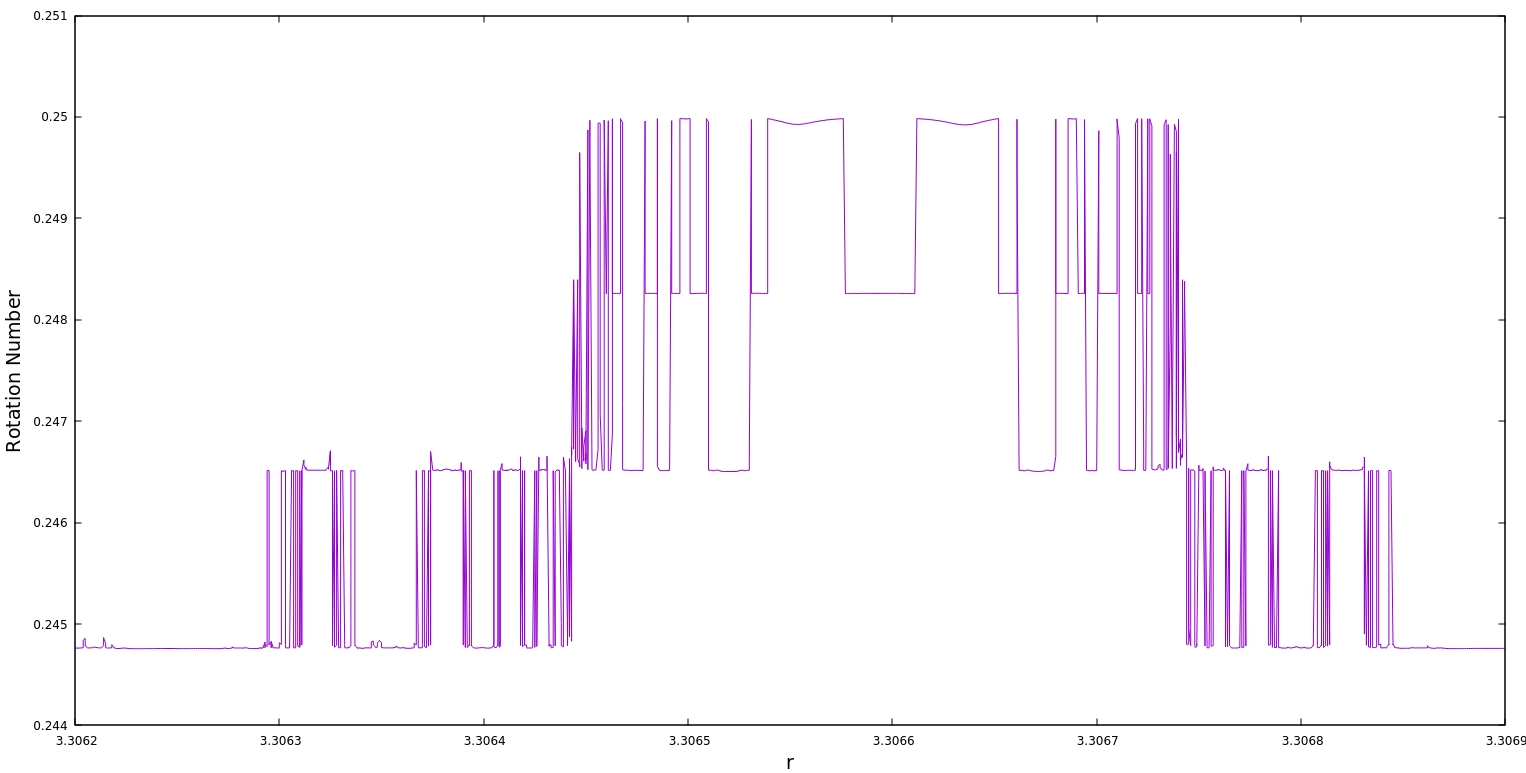}
    \caption{Detail of figure \ref{F:rotq09a}. As we change the scale of the rotation number in the chaotic interval, it is found a self similarity structure.}    
    \label{F:rotq09b}
\end{figure}

Figures \ref{F:rotq09a} and \ref{F:rotq09b} show the self similarity of the rotation number for $q = 0.9$ which is consistent with chaotic motion; at any scale we can find small constant rotation number (resonances) surrounded by rapid oscillations. The radial distance resolution of the figures \ref{F:rotq09a} and \ref{F:rotq09b} are of the order $10^{-5}$ and $10^{-7}$ respectively.

\begin{figure}[H]
    \centering
    \includegraphics[width=0.9\textwidth]{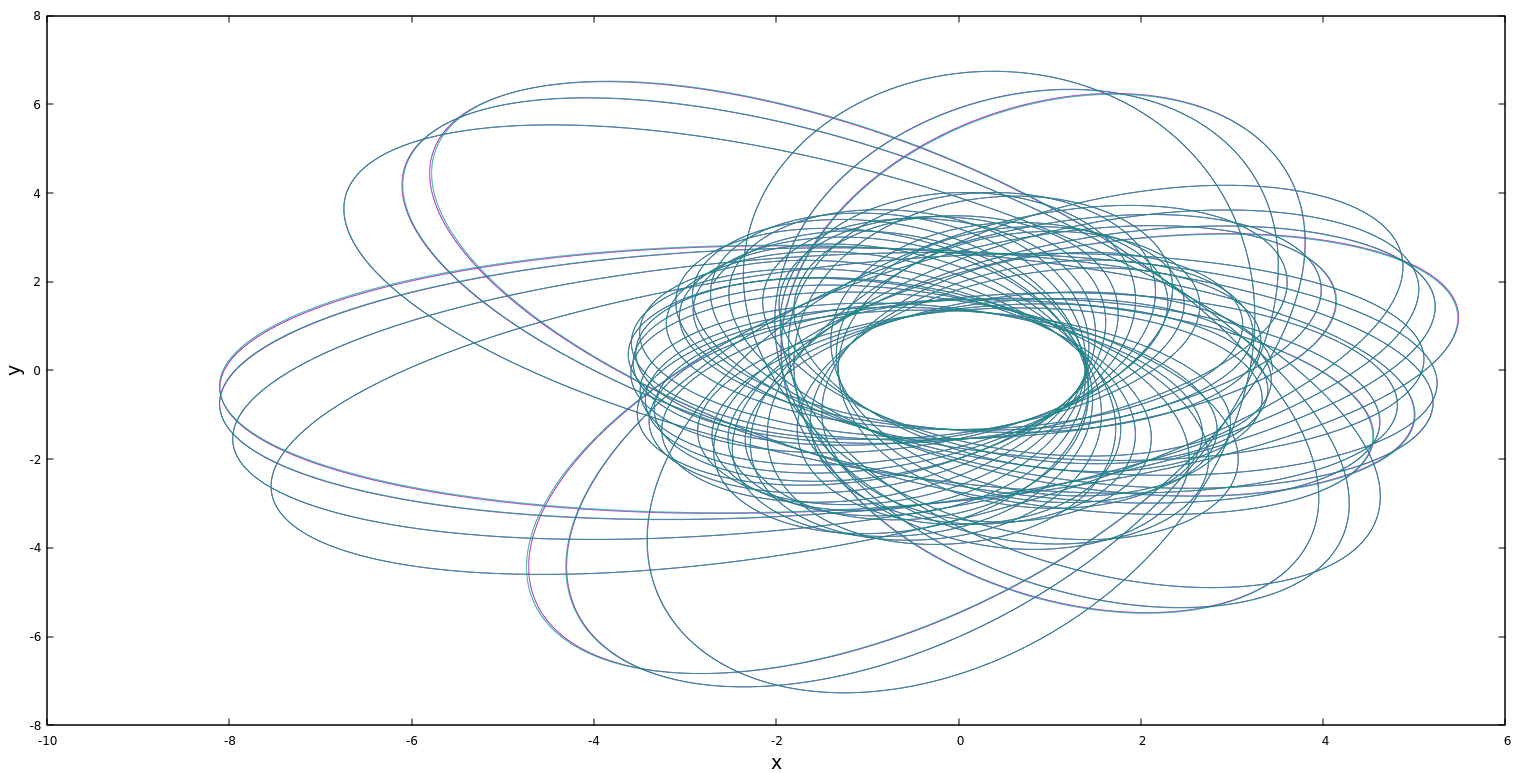}
    \caption{Projection on the equatorial plane of two orbits for $q = 0.9$ with a moderate simulation time and located in the chaotic region between $6/25$ and $9/37$ resonances shown in figure \ref{F:rotq09islas} with the same initial condition except for the initial radial position ($r_1 = 3.30011$ and $r_2 = 3.30012$). Both orbits remain \emph{stuck} during the travel, where they are practically indistinguishable.}    
    \label{F:confq09a}
\end{figure}

\begin{figure}[H]
    \centering
    \includegraphics[width=0.9\textwidth]{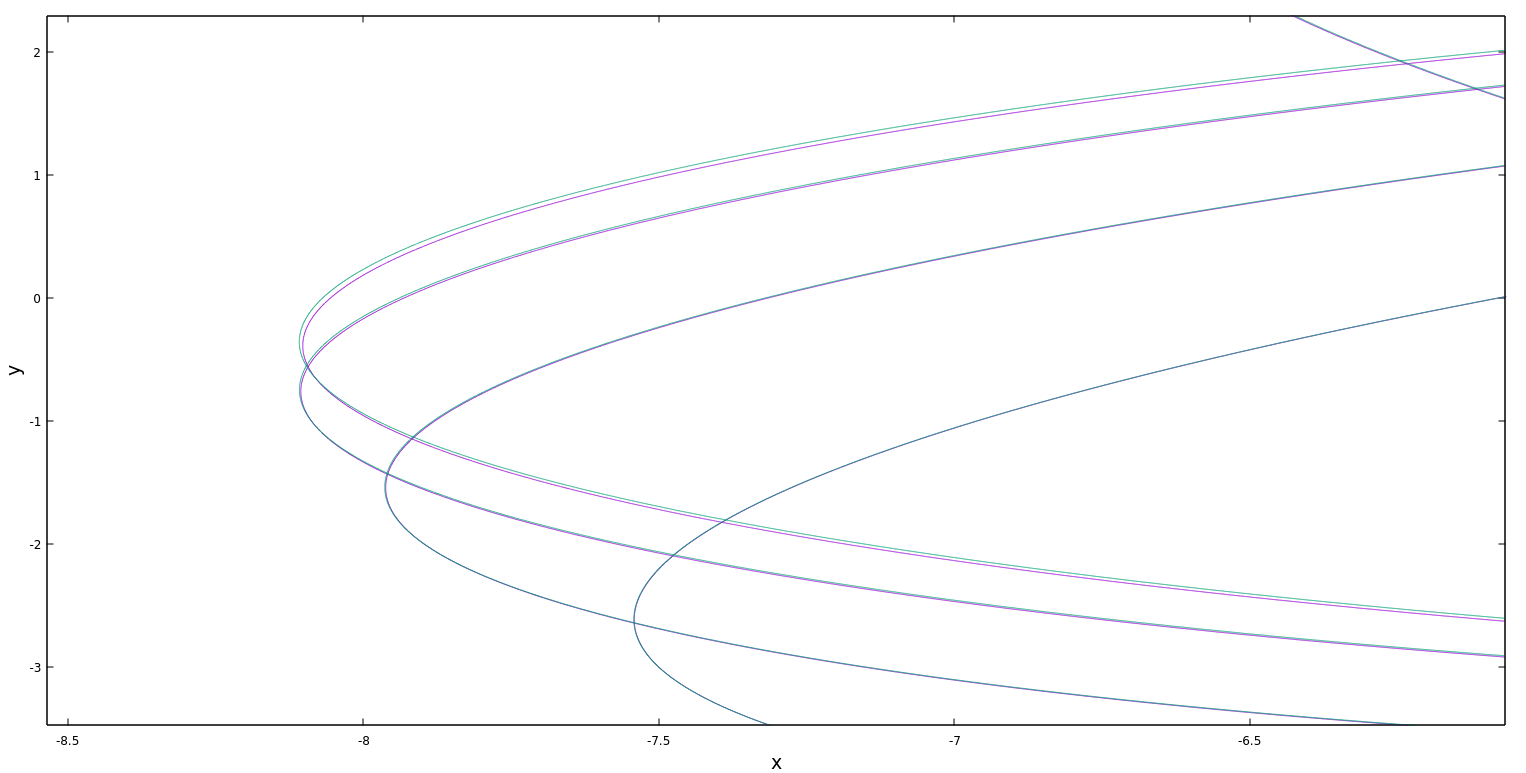}
    \caption{Detail of figure \ref{F:confq09a}. During their path the orbits are held together.}    
    \label{F:confq09a2}
\end{figure}

\begin{figure}[H]
    \centering
    \includegraphics[width=0.9\textwidth]{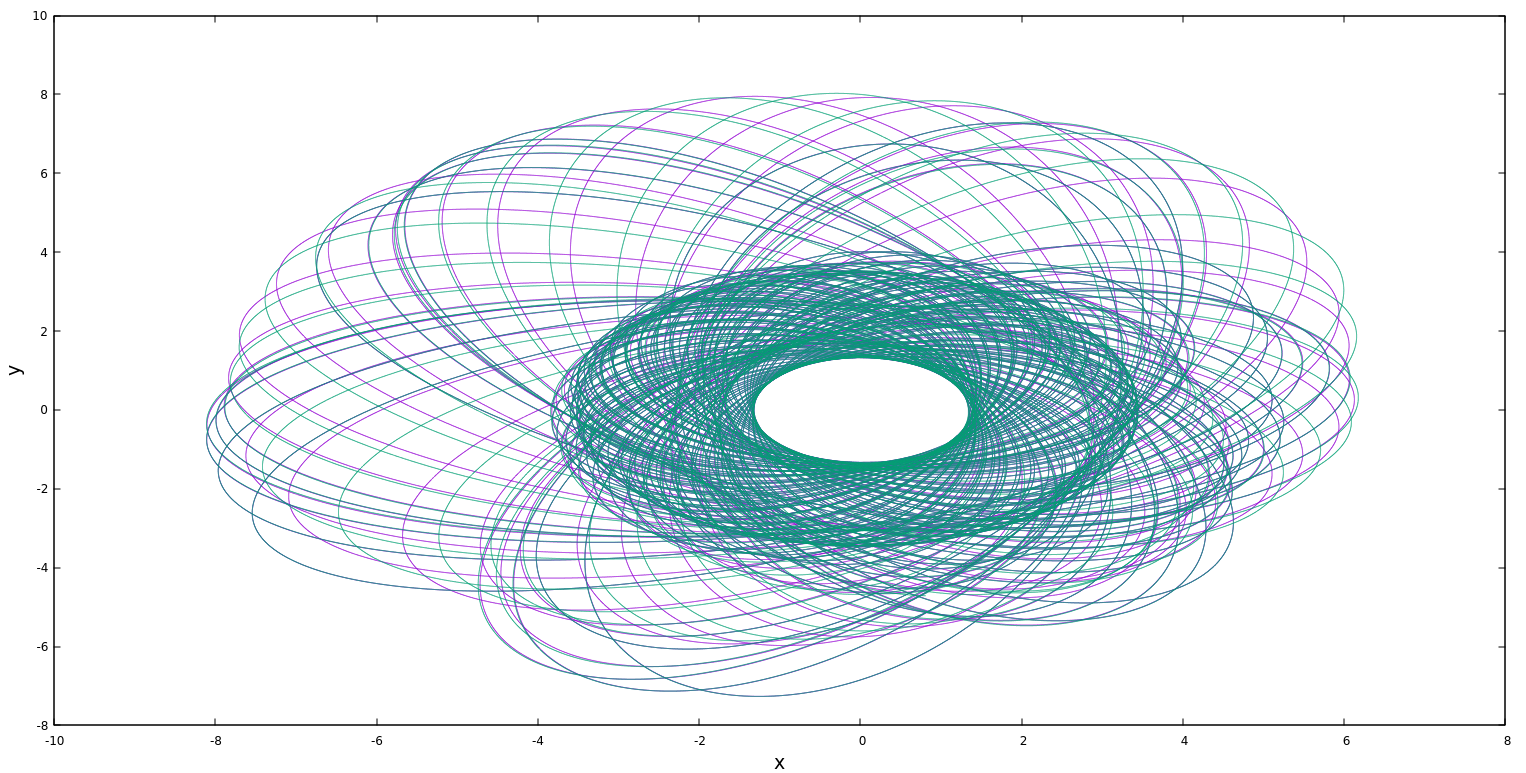}
    \caption{Prolongation of the trajectories of figure \ref{F:confq09a}, by increasing the simulation time the effect of staying in a chaotic region is visualized as both paths diverge from each other.}    
    \label{F:confq09b}
\end{figure}

\begin{figure}[H]
    \centering
    \includegraphics[width=0.9\textwidth]{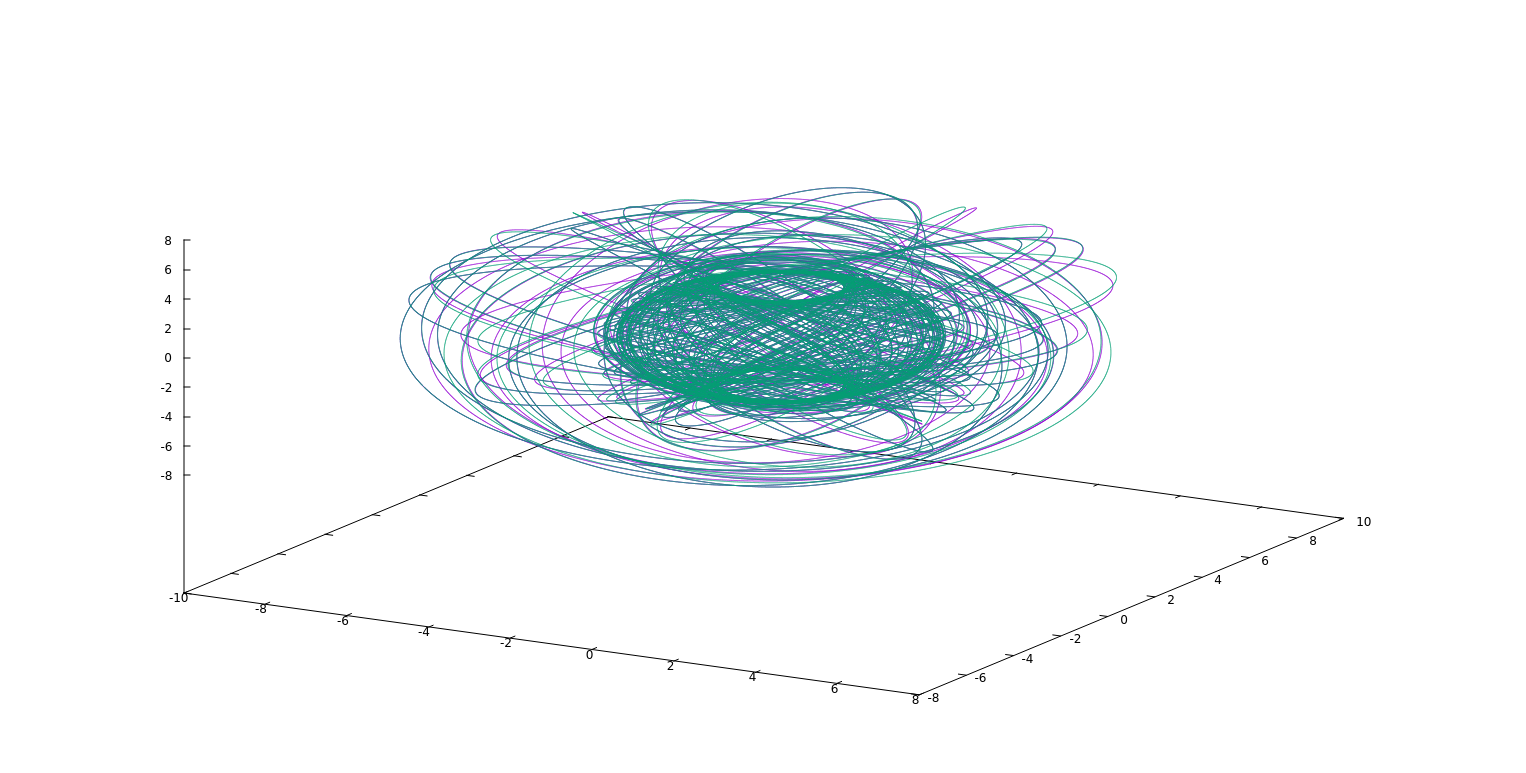}
    \caption{Configuration space of figure \ref{F:confq09b}. Both geodesics diverge when the simulation time is large enough.}    
    \label{F:confq09c}
\end{figure}

As previously mentioned, the rotation number identifies the chaotic regions and resonances as well. On the other hand, the stickiness makes it difficult to identify such orbits in the configuration space \cite{contopoulosgerakopoulos,contopoulosorderandchaos,countopouloslukes2014non,HowtoObserveNonKerr,LukesZpoyVoorhees,MoisesSantos} although it is a manifestation of chaotic motion because two nearby orbits will follow almost the same path for a long time. As shown in figures \ref{F:confq09a} and \ref{F:confq09a2} the two orbits are actually nearly indistinguishably.  Chaotic orbits will remain attached to higher order islands (satellites) until they diverge by the non linearities of the system \cite{LukesZpoyVoorhees}. Moreover, those orbits might fall to the event horizon, this behavior is present in the Manko-Novikov metric \cite{mankonovikov,contopoulosgerakopoulos} and the Zipoy-Vorheees metric \cite{LukesZpoyVoorhees}. However, figures \ref{F:confq09b} and \ref{F:confq09c} show the evolution of the same configuration space of figure \ref{F:confq09a} for a longer simulation time, as we see those orbits diverge but remain confined to the tori without falling to the event horizon nor they escape from the source gravitational field.   

\begin{figure}[H]
    \centering
    \includegraphics[width=0.9\textwidth]{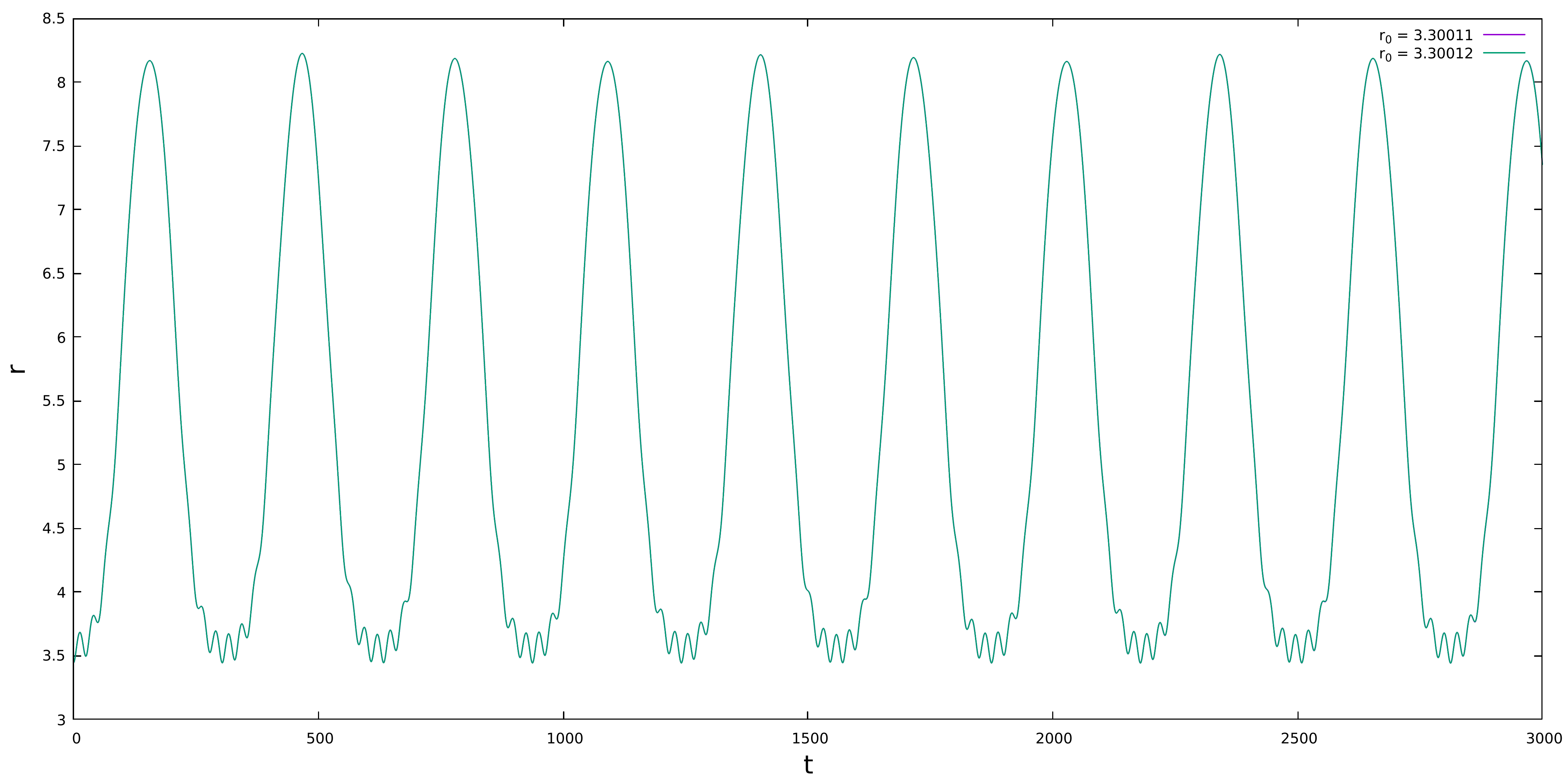}
    \caption{Radial evolution of the configuration space of figure \ref{F:confq09b}. Both geodesics are nearly indistinguishable as they started at $r = 3.30011$ and $r = 3.30012$.}    
    \label{F:r0time}
\end{figure}

On the other hand, it is possible to analyze the previous behaviour by looking the time evolution of the radial coordinate and the $\theta$-angular component. At the beginning of their trajectories, it is expected that the two geodesics have the same path because of the similar initial conditions as seen in figures \ref{F:r0time} and \ref{F:r0time}.

\begin{figure}[H]
    \centering
    \includegraphics[width=0.9\textwidth]{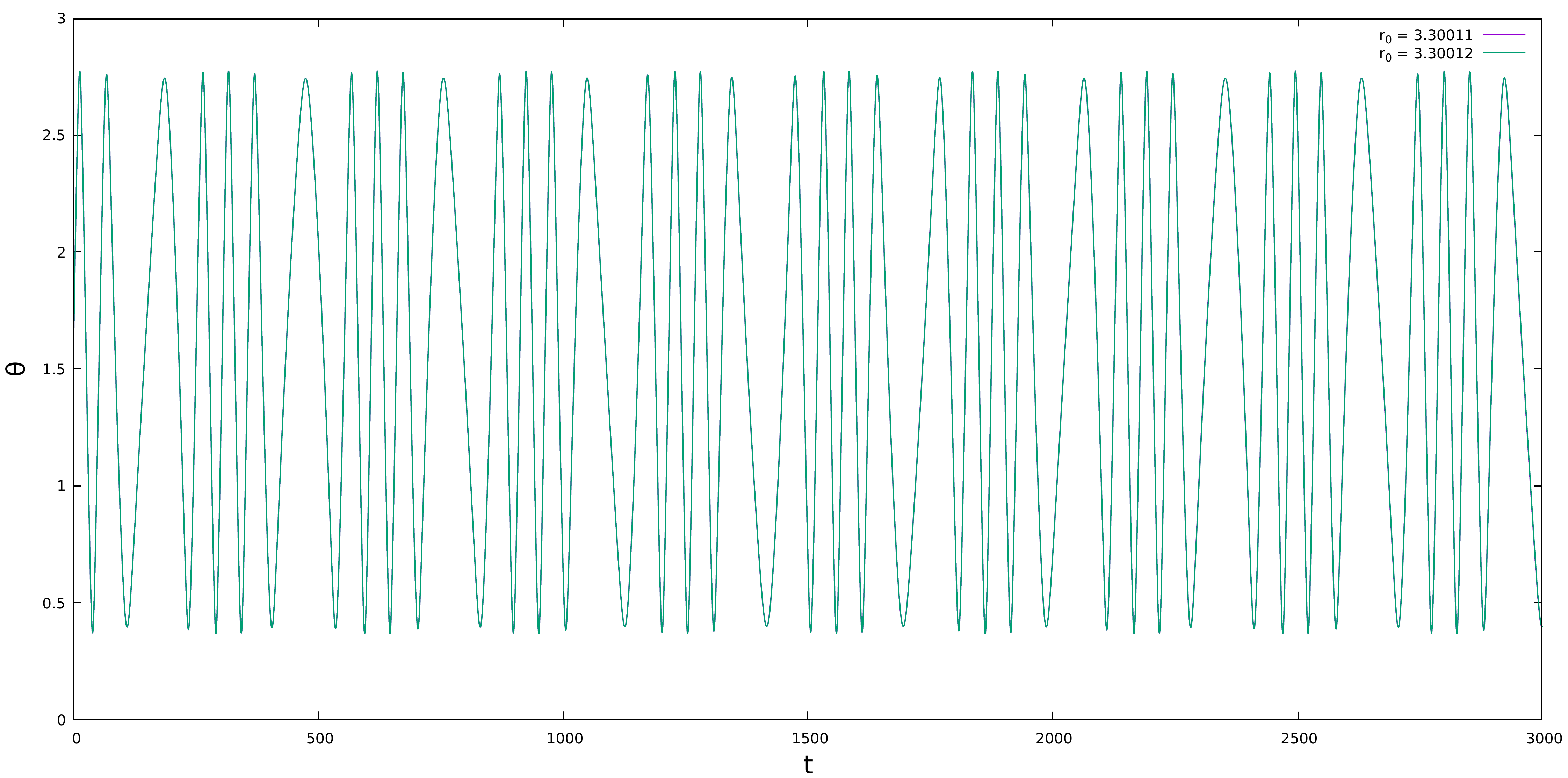}
    \caption{$\theta$-angle evolution of the configuration space of figure \ref{F:confq09b}. Both geodesics are nearly indistinguishable as they started at $r = 3.30011$ and $r = 3.30012$.}    
    \label{F:theta0time}
\end{figure}

\begin{figure}[H]
    \centering
    \includegraphics[width=0.9\textwidth]{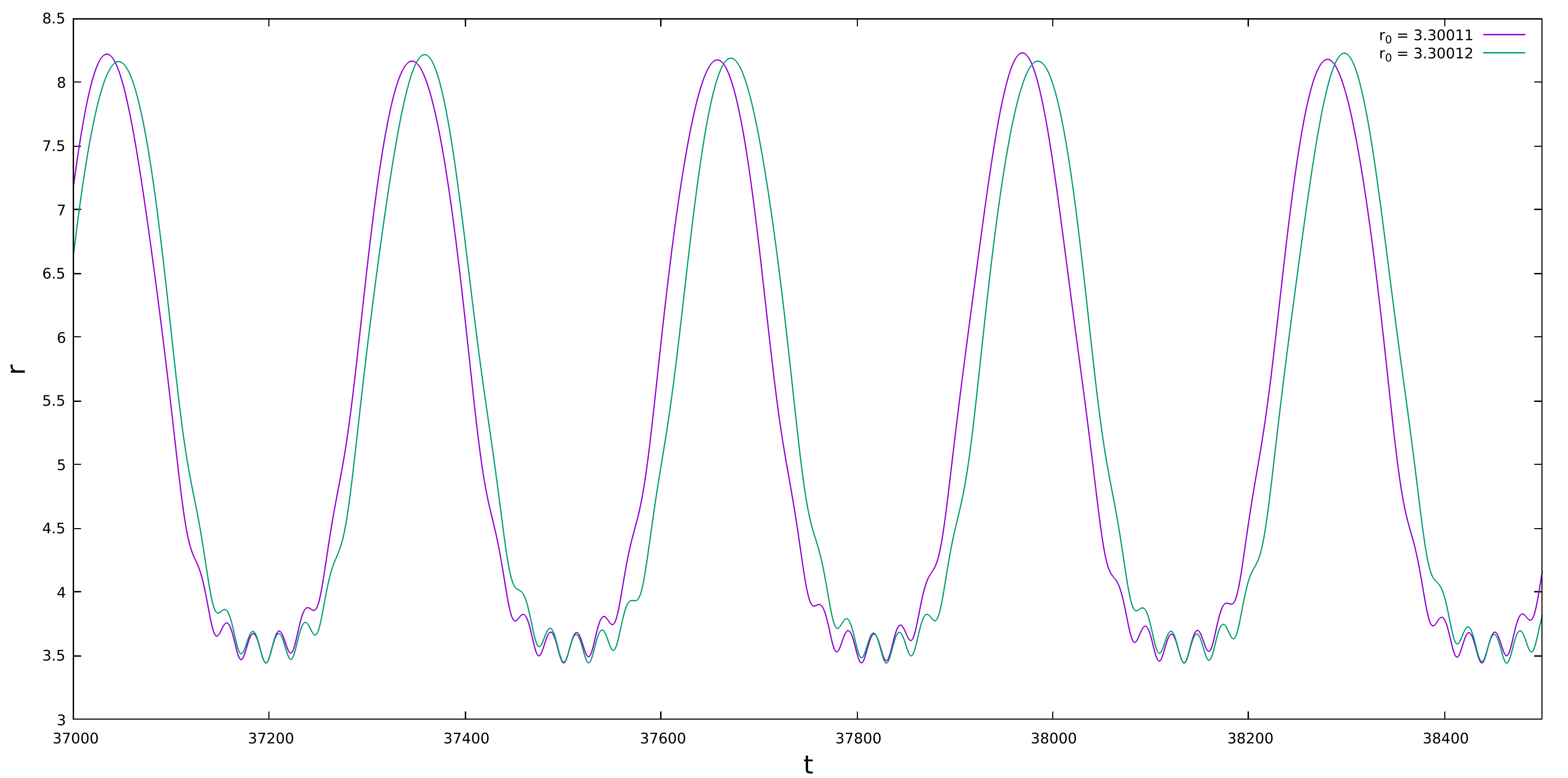}
    \caption{Radial evolution of the configuration space of figure \ref{F:confq09b}. They started at $r = 3.30011$ and $r = 3.30012$, but for a large time their behaviour diverge.}    
    \label{F:rlargetime}
\end{figure}

\begin{figure}[H]
    \centering
    \includegraphics[width=0.9\textwidth]{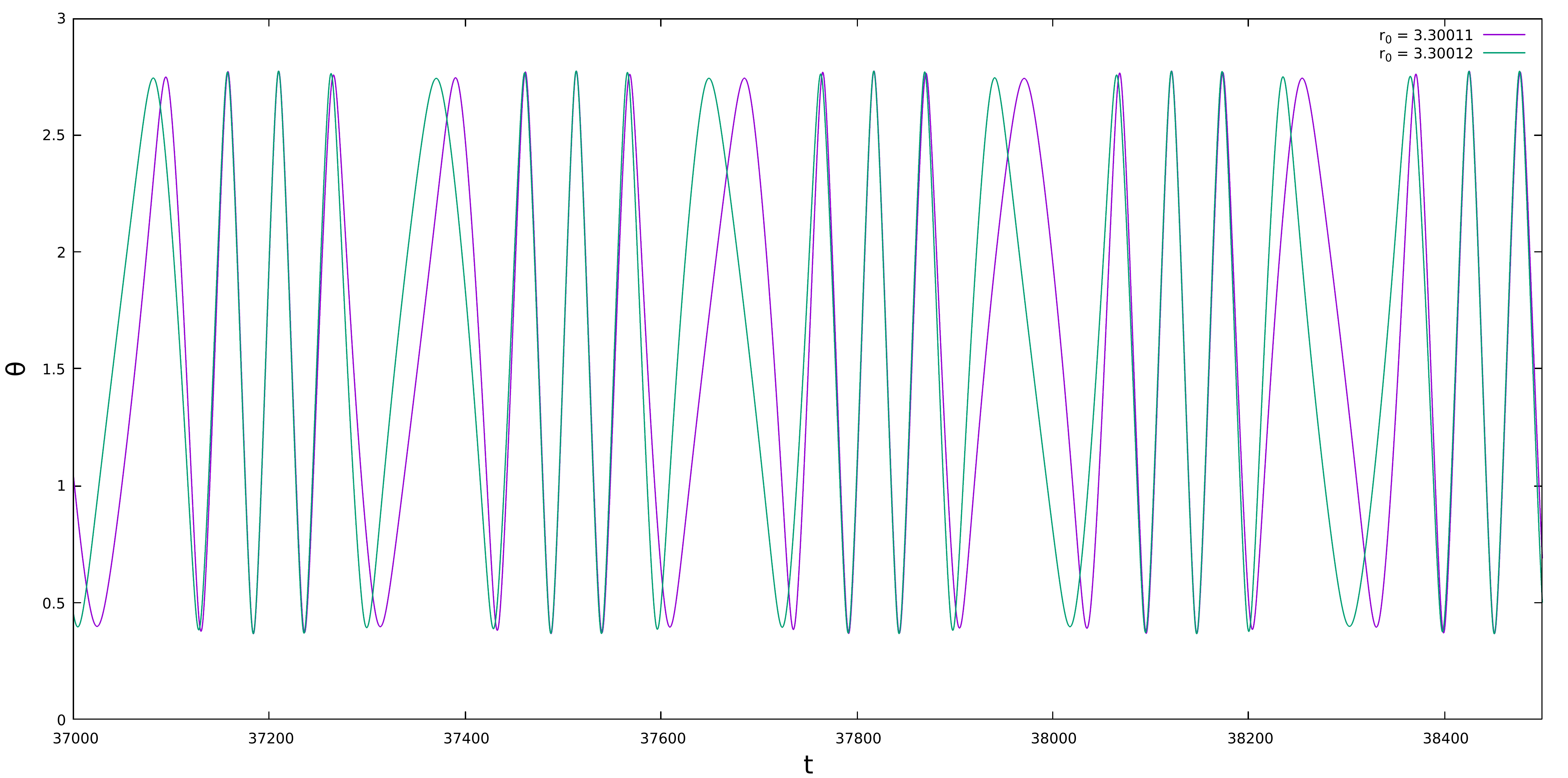}
    \caption{$\theta$-angle evolution of the configuration space of figure \ref{F:confq09b}. They started at $r = 3.30011$ and $r = 3.30012$, but for a large time their behaviour diverge.}    
    \label{F:thetalargetime}
\end{figure}

However, as the time increases the non linearities decouple the geodesics and both have independent paths. The magnitude of the variation in the rotation number in the chaotic region is relative small, therefore the previous geodesics will remain for a long time in nearly the same radial interval, but their evolution will be independent as shown in figures \ref{F:rlargetime} and \ref{F:thetalargetime}.

\begin{figure}[H]
    \centering
    \includegraphics[width=0.9\textwidth]{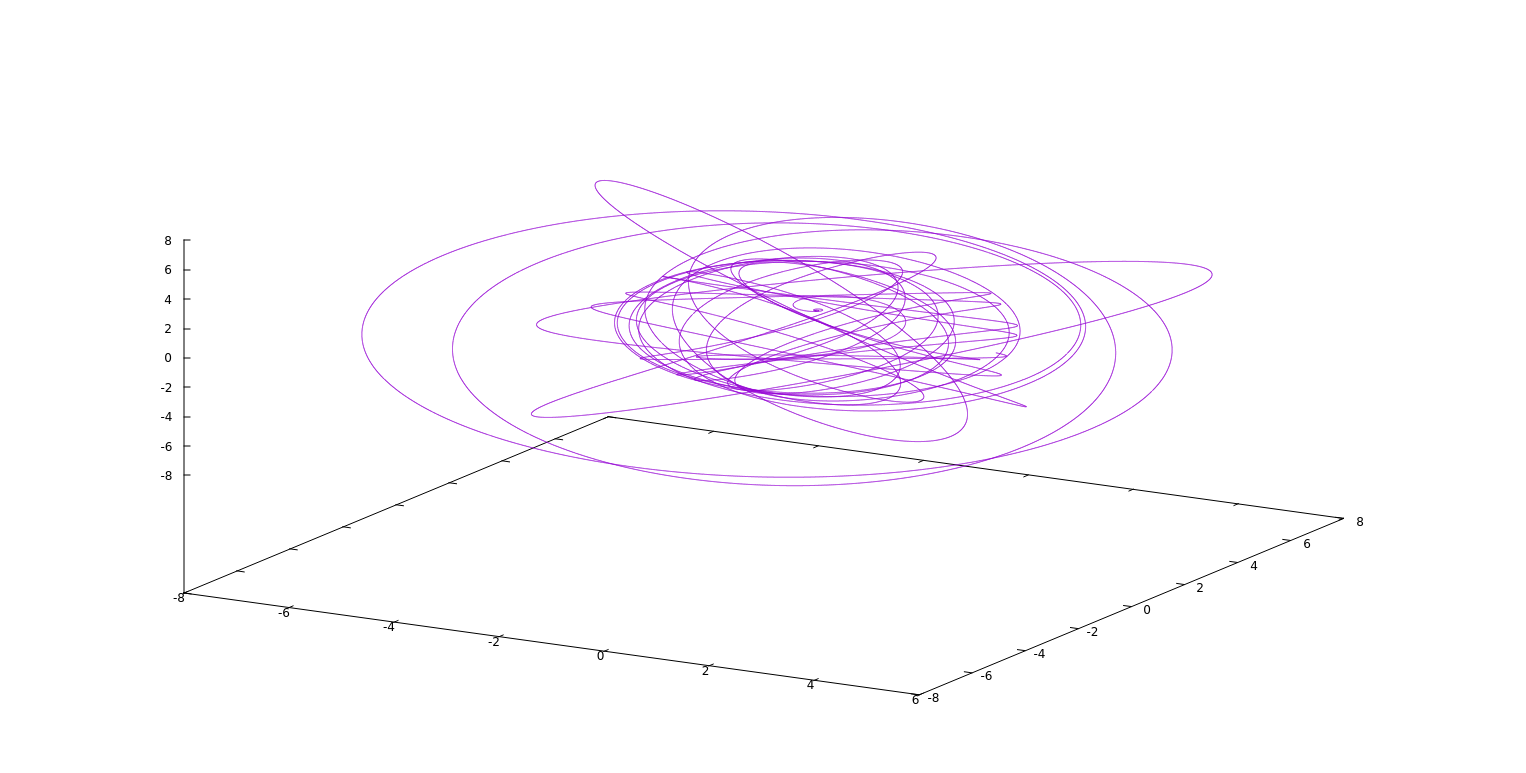}
    \caption{Trajectory of a chaotic geodesic for $q = 0.9$ located between the $2/9$ and $6/25$ resonances with initial condition $r = 3.29$. This orbit falls to the event horizon.}    
    \label{F:confq09d}
\end{figure}

Lastly, we found the same behavior described by \cite{contopoulosgerakopoulos, LukesZpoyVoorhees} for falling orbits. As shown in figure \ref{F:confq09d}, the orbit of the test particle follows a torus; but eventually it spirals toward the compact object.

\section{Conclusions and Final Remarks}

In this paper, the theory of dynamical systems together with hamiltonian mechanics was used to analyze the structure of the spacetime surrounding deformed compact objects. Taking the Kerr-like metric with mass quadrupole moment and the Hamilton's canonical equation, we have found the geodesic equations for the phase space. Later, we applied numerical integration of these equations to find the configuration space and construct the Poincaré section for a selected set of parameters and compare the evolution of the phase space by increasing the mass quadrupole moment. Finally the rotation number was constructed to analyze the behavior present in the Poincaré section.

As we have observed, the different structures in the Poincaré section implies that the system given by the Kerr-like metric is no-integrable, therefore it is required a numerical solution. The original Kerr system has torus stable orbits with no structures besides the main island of stability in the Poincaré section because of the existence of four constants of motion (energy $E$, angular momentum $L_z$, rest mass $\mu$ and Carter constant $\mathfrak{C}$). However, following the KAM theorem and Poincaré-Bendixon theorem, the tori will break setting $q \ne 0$, thus giving structures visible in the Poincare section. The remaining orbits only get deformations. This behavior of the Kerr-like system is due to the absence of the Carter constant, by setting $q \ne 0$ in the dynamical system loses its integrability property. Hence, the Poincaré sections are the ideal tools to determine the non linear evolution of the system.

In the region closest to the event horizon, due the intense gravity and the deformation of the compact object, several structures like chains of island, hyperbolic points, scattered points or higher order islands appear in the phase space. Those structures are identified with the Poincaré section or the rotation number along the $p_r = 0$ axis. We have identified these and incipient chaos at the border of the last stable orbit even for very low deformation ($q = 0,1$). Using the rotation number for several values of $q$, different chaotic regions were identified. Moreover, it was found that $2/9, \, 2/7$ and $2/5$ resonances survive as the mass quadrupolar moment changes. The $2/9$ resonance with high $q$ will be completely surrounded by higher order islands and a sea of chaotic geodesics. Additionally, the chaotic regions and the stable regions are separated by a hyperbolic points; most notably for $q = 0.9$ the hyperbolic point at $r = 3.308$ abruptly ends the chaotic zone as shown in figure \ref{F:rotq09}. Lastly, the rotation number shows self similarity in the chaotic intervals.

The stickiness phenomenon is extremely important to characterize the behavior of the chaotic geodesics, as they remain attached to the stable ones over a long period of time until they diverge. Some orbits remain attached to higher order islands and others flow towards the event horizon. Both behaviors were studied for $q = 0.9$.

Contopoulos, et al. \cite{contopoulosgerakopoulos, LukesZpoyVoorhees} found several structures which are not present in the Kerr-like system. Particularly, \cite{contopoulosgerakopoulos} islands of stability extremely near the singularity, which are not present in this study. We conclude that the main behaviour surrounding the main island of stability could be found for other metrics due the breaking of spherical symmetry, but certain features depends directly on the metric used to describe the spacetime source. 

As an additional remark, the recent advances in direct observation of accretion disk of the black holes, stars orbiting massive compact objects and the detection gravitational waves could provide information about the shape of the gravitational source. Moreover, the behavior of the spectra of gravitational waves is directly related with the phase space, and the study of different metrics and their respective chaotic regions will be important in order to understand the evolution of the system.



\begin{thebibliography}{99}

\bibitem{contopoulosorderandchaos}
{Contopoulos, G.}
{Order and Chaos in Dynamical Astronomy}, {Springer}, {2004},
  
\bibitem{contopoulosgerakopoulos}
{Contopoulos, G., Lukes-Gerakopoulos, G., Apostolatos, T. A.}
{Orbits in a Non-{K}err Dynamical System},
{International Journal of Bifurcation and Chaos}, {21}({08}), {2261--2277}, {2011}. {https://doi.org/10.1142/S0218127411029768}

\bibitem{frutosmetric}
 {Frutos-Alfaro, F.}
{Approximate {K}err-Like Metric with Quadrupole}, 
{International Journal of Astronomy and Astrophysics}, {6}, {334--345}, {2016}. \\
{https://doi.org/10.4236/ijaa.2016.63028}

\bibitem{frutos2}
{Frutos-Alfaro, F.}
{Approximate Spacetime for Neutron Stars}
{General Relativity and Gravitation}, {51}, {46}, {2019}. {https://doi.org/10.1007/s10714-019-2530-5}

\bibitem{goldstein}
{Goldstein, H. and Poole, C., Safko, J.} 
{Classical {M}echanics}, {Addison Wesley}, {San Francisco}, {2000}.

\bibitem{grossman2012harmonic} 
{Grossman, R., Levin, J., Perez-Giz, G.}
{Harmonic {S}tructure of {G}eneric {K}err {O}rbits}, 
{Physical Review. D}, {85}({2}), {023012}, {2012}. \\ {https://doi.org/10.1103/PhysRevD.85.023012}

\bibitem{countopouloslukes2014non}
{Lukes-Gerakopoulos, G., Contopoulos, G., Apostolatos, T. A.}
{Non-Linear Effects in {N}on-{K}err Spacetimes}, 
{Bi{\u c}\'ak J., Ledvinka T. (eds)} {Relativity and Gravitation}, 
{Springer Proceedings in Physics}, {157}, {2014}.
{https://doi.org/10.1007/978-3-319-06761-2{\_}16}

\bibitem{mankonovikov}
{Manko, V. S., Novikov, I. D.}
{Generalizations of the Kerr and Kerr-Newman Metrics Possessing an Arbitrary Set of Mass-Multipole Moments}, {Classical and Quantum Gravity}, {9}({11}), {2477--2487}, {1992}.
{https://doi.org/10.1088/0264-9381/9/11/013}

\bibitem{gravitationmisnerthorne}
{Misner, C. W., Thorne, K. S., Wheeler, J. A.}
{Gravitation}, {Freeman}, {New York}, {1973}. 

\bibitem{weinberg}
{Weinberg, S.}
{Gravitation and Cosmology: Principles and Applications of the General Theory of Relativity}, {John Wiley and Sons, Inc.}, {New York}, {1972}.

\bibitem{LukesZpoyVoorhees}
{Lukes-Gerakopoulos, Georgios.}
{Nonintegrability of the Zipoy-Voorhees metric}, {Phys. Rev. D},{86}, {2012}. {https://doi.org/10.1103/PhysRevD.86.044013}

\bibitem{HowtoObserveNonKerr}
{Apostolatos T. A., Lukes-Gerakopoulos G., Contopoulos G.}, {How to observe a non-Kerr spacetime using gravitational waves}, {Phys Rev Lett.}, {103(11):111101.}, {2009}. {https://doi.org/10.1103/PhysRevLett.103.111101}

\bibitem{AngulardynamicsVoglis}
{Voglis, N., Efthymiopoulos, C.}, {Angular dynamical spectra. A new method for determining frequencies, weak chaos and cantori.}, {Journal of Physics A: Mathematical and General}, {31}, {1998}. {https://doi.org/10.1088/0305-4470/31/12/015}

\bibitem{FrequencyFourier}
{Laskar, J.}, {Frequency Analysis of a Dynamical System}, {Celestial Mechanics and Dynamical Astronomy.}, {56}, {1993}. {https://doi.org/10.1007/BF00699731}

\bibitem{FrequencyFourierLaskar}
{Papaphilippou, Y., Laskar, J.}, {Frequency map analysis and global dynamics in a galactic potential with two degrees of freedom.}, {Astronomy and Astrophysics.}, {307}, {1996}. {https://ui.adsabs.harvard.edu/abs/1996A\&A...307..427P}

\bibitem{MoisesSantos}
{Santos , M., Mugnaine, M., Szezech, J., Batista, A., Caldas, I., Viana, R.}, {Using rotation number to detect sticky orbits
in Hamiltonian systems.}, {AIP Publishing}, {Chaos 29, 043125}, {2019}. {https://doi.org/10.1063/1.5078533}

\bibitem{BrinkResonant}
{Brink, J., Geyer, M., Hinderer, T.}, {The Astrophysics of Resonant Orbits in the Kerr Metric.}, {Phys. Rev. D.}, {91}, {083001}, {2015}. {https://doi.org/10.1103/PhysRevD.91.083001}

\bibitem{ChaoticShadow}
{Wang, M., Chen, S., Jing, J.}, {Chaotic shadow of a non-Kerr rotating compact object with quadrupole mass moment.}, {Phys. Rev. D}, {98}, {104040}, {2018}. {https://doi.org/10.1103/PhysRevD.98.104040}

\bibitem{GrowthResonances}
{Zelenka, O., Lukes-Gerakopoulos, G.,  Witzany, V., Kopáček, O.}, {Growth of resonances and chaos for a spinning test particle in the Schwarzschild background}, {Phys. Rev. D}, {American Physical Society}, {101}, {024037}, {2020}. {https://doi.org/10.1103/PhysRevD.101.024037}

\bibitem{TestingSpacetime}
{Destounis, K., Suvorov, A. G., Kokkotas, K. D.}, {Testing spacetime symmetry through gravitational waves from extreme-mass-ratio inspirals}, {Phys. Rev. D}, {102},  {064041}, {2020}. {https://doi.org/10.1103/PhysRevD.102.064041}

\end{thebibliography}

\end{document}